# The Geology of Pluto and Charon Through the Eyes of New Horizons

Jeffrey M. Moore[1]*, William B. McKinnon[2], John R. Spencer[3], Alan D. Howard[4], Paul M. Schenk[5], Ross A. Beyer[6,1], Francis Nimmo[7], Kelsi N. Singer[3], Orkan M. Umurhan[1], Oliver L. White[1], S. Alan Stern[3], Kimberly Ennico[1], Cathy B. Olkin[3], Harold A. Weaver[8], Leslie A. Young[3], Richard P. Binzel[9], Marc W. Buie[3], Bonnie J. Buratti[10], Andrew F. Cheng[8], Dale P. Cruikshank[1], Will M. Grundy[11], Ivan R. Linscott[12], Harold J. Reitsema[3], Dennis C. Reuter[13], Mark R. Showalter[6], Veronica J. Bray[14], Carrie L. Chavez[6,1], Carly J. A. Howett[3], Tod R. Lauer[15], Carey M. Lisse[8], Alex Harrison Parker[3], S. B. Porter[3], Simon J. Robbins[3], Kirby Runyon[8], Ted Stryk[16], Henry B. Throop[17], Constantine C. C. Tsang[3], Anne J. Verbiscer[18], Amanda M. Zangari[3], Andrew L. Chaikin[19], Don E. Wilhelms[20]

[1]National Aeronautics and Space Administration (NASA) Ames Research Center, Space Science Division, Moffett Field, CA 94035, USA.
[2]Department of Earth and Planetary Sciences, Washington University, St. Louis, MO 63130, USA.
[3]Southwest Research Institute, Boulder, CO 80302, USA.
[4]Department of Environmental Sciences, University of Virginia, Charlottesville, VA 22904, USA.
[5]Lunar and Planetary Institute, Houston, TX 77058, USA.
[6]The SETI Institute, Mountain View, CA 94043, USA.
[7]University of California, Santa Cruz, CA 95064, USA.
[8]Johns Hopkins University Applied Physics Laboratory, Laurel, MD 20723, USA.
[9]Massachusetts Institute of Technology, Cambridge, MA 02139, USA.
[10]NASA Jet Propulsion Laboratory, La Cañada Flintridge, CA 91011, USA.
[11]Lowell Observatory, Flagstaff, AZ 86001, USA.
[12]Stanford University, Stanford, CA 94305, USA.
[13]NASA Goddard Space Flight Center, Greenbelt, MD 20771, USA.
[14]University of Arizona, Tucson, AZ 85721, USA.
[15]National Optical Astronomy Observatory, Tucson, AZ 85719, USA.
[16]Roane State Community College, Oak Ridge, TN 37830, USA.
[17] Planetary Science Institute, Tucson, AZ 85719, USA.
[18]Department of Astronomy, University of Virginia, Charlottesville, VA 22904, USA.
[19] Independent Science Writer, Arlington, VT, USA.
[20]U.S. Geological Society, Retired, Menlo Park, CA, USA.

**Abstract:** NASA's New Horizons spacecraft has revealed the complex geology of Pluto and Charon. Pluto's encounter hemisphere shows ongoing surface geological activity centered on a vast basin containing a thick layer of volatile ices that appears to be involved in convection and advection, with a crater retention age no greater than ~10 Ma. Surrounding terrains show active glacial flow, apparent transport and rotation of large buoyant water-ice crustal blocks, and pitting, likely by sublimation erosion and/or collapse. More enigmatic features include tall mounds with central depressions that are conceivably cryovolcanic, and ridges with complex bladed textures. Pluto also has ancient cratered terrains up to ~4 Ga old that are extensionally fractured and extensively mantled and perhaps eroded by glacial or other processes. Charon does not appear to be currently active, but experienced major extensional tectonism and resurfacing (probably cryovolcanic) nearly 4 billion years ago. Impact crater populations on Pluto and Charon are not consistent with the steepest proposed impactor size-frequency distributions proposed for the Kuiper belt.

**One Sentence Summary:** NASA's New Horizons spacecraft has revealed surprisingly complex geology on the surfaces of Pluto and Charon, with probable ongoing activity on Pluto.

## Introduction

We present a preliminary geological examination of Pluto and Charon based on images and other data collected by NASA's New Horizons spacecraft during its flyby of these worlds on 14 July 2015. The two camera systems pertinent to geological investigations are the wide-angle color Multispectral Visible Imaging Camera (MVIC) and the narrow-angle panchromatic LOng Range Reconnaissance Imager (LORRI) (*1*). This paper focuses on the portions of the illuminated surfaces seen near closest approach at better than 1 km/pixel resolutions, centered on 180° longitude for Pluto and 0° longitude for Charon (*2, 3*). All topographic measurements were obtained using stereo photogrammetric techniques, supplemented by shadow and limb measurements (*4*). An acronym list is provided in the Supplementary Online Material (SOM). All feature names used in this paper are informal, and the locations of named features on Pluto and Charon are shown in Figs. S1 and S2 respectively; terrain locations are indicated in Fig. S3.

## Pluto

Pluto's surface exhibits an astonishing variety of landscapes (Fig. 1A). Broadly, the encounter hemisphere (EH) contains several regional provinces: (1) the ~1000 km wide uncratered plain, Sputnik Planum (SP), centered on the EH; (2) arcuate, rugged-to-mountainous regions surrounding SP on three sides; (3) mantled and eroded plains at higher latitudes; and (4) a heterogeneous surface west of SP containing plains with various degrees of crater density and surface texture, scarps (both erosional and tectonic), troughs (graben), and patches of rugged cratered terrain.

***Sputnik Planum and environs***

This ~870,000 km$^2$ oval-shaped unit of high-albedo plains, centered at ~20°N, 175°E, is likely a massive unit of volatile ices (solid $N_2$, CO and $CH_4$) (*6*), the level of which is 3-4 km below the surrounding uplands. The central and northern regions of SP display a distinct cellular pattern (*6*), which varies in appearance across the planum. In the bright central portion (Fig. S4A), the cells are bounded by shallow troughs up to 100 m deep (*6*); the centers of at least some cells are elevated by ~50 m relative to their edges, though some apparently have less relief. The southern region and eastern margin of SP do not display cellular morphology, instead showing featureless plains and dense concentrations of pits, themselves reaching a few kilometers across (Fig. S4D). Details of the different morphologies encountered within SP are described in the SOM.

No impact craters have been confirmed on SP in contiguous mapping coverage at 390 m/pixel scale. Following the arguments in (*6*), the crater retention age of SP is very young ($\lesssim$10 Ma), and is discussed in the SOM. Such geologically recent resurfacing and/or topographic relaxation is consistent with the weak rheology of $N_2$-dominated ices (*7, 8*), and with the interpretation of cells as expressions of potentially active solid-state convection in a thick layer of such ices (SOM).

A discontinuous chain of mountains, consisting of discrete angular blocks with apparently random orientations and sizes up to 40 km across and 5 km high, extends for hundreds of km along the west margin of SP. Those in the south are often separated by embaying materials, whereas those in the north, particularly the northernmost al-Idrisi Montes (AIM, Fig. 2), have minimal separation. At AIM, blocks are closely packed, and many blocks have flat or gently sloping upper surfaces with linear textures similar to some of the surrounding highland terrain, suggesting breakup of a preexisting surface.

The northern inter-block material has a distinctive reddish color (Fig. 2A), contains many smaller blocks, and is slightly elevated relative to SP; similar terrain surrounds some of the mountains to the south. The AIM region contains two depressions floored largely by this finer, inter-block material and small blocks (‘c’ in Fig. 2), and another occupied by a small plain with similar texture and color to SP (‘e’ in Fig. 2). An inward-facing terrace surrounds this depression (‘d’ in Fig. 2), suggesting an earlier, higher level of plains material.

It was argued by (*6*) that the steep slopes and high elevations of the mountain blocks require a water-ice-based composition; this has now been confirmed (*9*). Like the angular blocks in europan chaos, Pluto’s mountain blocks appear to consist of fragments of pre-existing ice crust that have been detached by fracturing, transported, and rotated. The exclusive location of this chaotic, blocky mountainous terrain on the margins of SP (Fig. S3), which evidently contains a significant thickness of low-viscosity ices, makes it plausible that these ices play a role in the disruption of Pluto’s crust.

$H_2O$ ice is buoyant with respect to $N_2$ and CO ice, but not $CH_4$ ice, and blocks of $H_2O$ ice embedded or buried in solid $N_2$ and/or CO will tend to rise isostatically. Small blocks can be carried along by convective or advective motions, essentially as icebergs, and large blocks may be undermined, shifted, and rotated. If the solid $N_2$/CO ice is sufficiently deep, then several of the smaller mountains may be floating within the plains, although the reliefs of the largest mountains (2-3 km), which skirt the western margin of SP, implies that their keels are likely ‘grounded’ on the basement (SOM). Why mountainous terrains within SP are limited to its western margin is unknown.

***Pits, blades, plains, and glaciers east of SP***

An intricate, high-albedo, 500 km wide landscape of pitted uplands and smooth plains, bordered by lower albedo bladed terrain, forms most of the eastern portion of Tombaugh Regio (TR).

*Pitted uplands:* The dominant features are pits ('a' in Fig. 3A), most of which are a few km across, but some exceed 25 km, locally intersecting to form long, linear troughs. Based upon preliminary topography, pits average ~1 km deep. The crests of the pits define an undulating upland surface 2 to 4 km above SP. In parts of the uplands the pitting is organized into distinct NE-SW trending ridge-and-trough terrain with ~5 km crest-to-crest spacing. The sidewalls of the pits typically slope up to 30º, suggesting rigid material underlies the thin, bright surface layer.

*Bladed terrain:* The pitted uplands transition northeastward to several broad (~100 km wide) swells named Tartarus Dorsa (TD), whose flanks and crests are covered with numerous roughly aligned blade-like ridges oriented ~N-S (Fig. 3B). Individual ridges are typically several hundred meters high, and are spaced 5 to 10 km crest to crest, separated by V-shaped valleys with slopes of ~20°. Many ridges merge at acute angles to form Y-shape junctions in plan view. Along the west flank of TD are a number of triangular to rectangular facets of the plains ramping upward toward the east.

*Smooth plains:* Nearly level expanses of smooth plains up to 50 km across occur at relative low points in the pitted uplands as well as elevated terraces adjacent to SP ('b' in Fig. 3A). They are generally smooth at 300 m/pixel resolution, but locally collections of km-scale hills extend above the plains, probably as protrusions or embedded fragments of the pitted terrain material.

The smoothness of the level plains suggests that they are composed of deformable ices, probably similar in composition to SP.

*Glaciers:* At a few locations along the SP/pitted uplands boundary, smooth materials connect with SP along the floors of troughs 1.5 to 6 km wide ('b' in Fig. 3A). High-phase imaging of the southernmost of these systems reveals conspicuous medial flow lines within the troughs extending onto SP, with the ice in the troughs sloping 2-3° over more than 50 km (SOM). This pattern implies glacial-like flow of the plains material into SP, perhaps analogous to ice streams at the margins of terrestrial ice sheets. At present it is unresolved whether the flowing ice carved the troughs.

*Origin and Evolution of Pitted Uplands, Bladed Terrain and Smooth Plains:* Both the pitted uplands and bladed terrain may be remnants of a formerly continuous deposit degraded either by sublimation (forming features analogous to those of degraded terrestrial snow or ice fields - penitentes and sun-cups - but much larger), or through undermining and collapse, possibly through melting at depth. An additional possibility is growth of ridges through preferential deposition of volatiles on ridge crests, analogous to pinnacle formation on Callisto (*3*). The preferential orientation of troughs and ridges in both terrains suggests an origin influenced by solar illumination direction and/or atmospheric circulation. In the case of the bladed terrain, if the material forming it was exposed through upwarping and erosion, it may have been a once-buried layer. The high albedo of the pitted uplands suggests condensation of volatiles sublimated and transported from SP (the pits that are prevalent on south SP may form through sublimation of $N_2$ ice, see SOM); these volatiles may accumulate to form the smooth plains.

***Upland terrains: Washboard and dissected terrains***

The uplands north and northwest of SP contain a variety of morphologies, notably including expanses of parallel ridges and troughs we call washboard terrain, and dissected terrain locally organized into valley networks. Fretted terrain and eroded mantles are discussed in the SOM.

*Washboard terrain:* Many flat expanses in this region feature parallel ridges and grooves with a crest-to-crest wavelength of about 1 km (Fig. 4A). The ridges retain a consistent NE-SW orientation, even where developed on the interior floors of craters. The albedo of washboard surfaces matches that of nearby ungrooved terrain, and underlying terrain features remain visible where grooved. These observations suggest that washboarding is a superficial modification, either by erosion of the underlying surface or, alternatively, as part of a thin regional deposit. The grooving is superimposed on higher-relief topographic features such as ridges, craters and dissected terrain. Occasional 1-2 km diameter craters are superposed on the washboarding.

*Dissected terrain:* Terrains dissected by valleys are common on the EH, including fluted, dendritic, plateau, alpine and mountainous variants (Fig. 4B). Two of these types occur widely. Fluted terrain containing troughs 15-20 km across with up to 2 km relative relief that are eroded into broad hills constitute one of these. The troughs or flutes are regularly spaced at 3-4 km and are oriented downhill with slopes up to 20º. The interior walls of some craters are similarly fluted. These troughs terminate abruptly in depressions or crater floors without evidence of deposition. Similarly spaced dendritic valley networks are another type of dissected terrain. The networks generally terminate in broad depressions. Dissected terrain appears to post-date and modify the larger upland craters. The other, less common, styles of dissection are described in the SOM.

*Origins of these terrains:* The mechanisms regulating the characteristic scale and groove orientation of washboarding remain uncertain. In the dissected terrain, both the fluted terrain and the dendritic valley networks probably result from advective processes, most likely flow of nitrogen-rich ice, possibly accompanied by basal melting (SOM). The spatial variation in morphology of the valley networks is likely to be a response to local topographic setting, substrate properties, latitudinal variations in insolation, and variation in depths and durations of $N_2$ ice accumulation.

### *Cthulhu Regio (CR)*

CR is a large dark area that covers a swath from ~15°N to ~20°S (Fig. S1), bordering TR at 160°E, and stretching westward almost half way around the planet to 20°E. Eastern CR is not a distinct physiographic province, but instead a region of dark mantling thin enough to preserve underlying topography, superimposed upon various geological terrains, including dendritic valleys, craters, fossae (long, narrow troughs), and retreating scarps. The dark coating is likely the result of atmospheric tholin deposition (*9*). CR contains striking correlations between color/albedo and topography: bright material is correlated with high elevations in some areas and with north-facing slopes in others. This may result in part from insolation-dependent deposition of the bright material on the dark landscape. Other western low latitude terrains are discussed in the SOM.

### *Large mounds with central depressions*

To the southwest of Norgay Montes (Figs. S1C and S3) are two broad quasi-circular mounds (Fig. 5). The northernmost (Wright Mons, WM) is 3-4 km-high and ~150 km across. At its summit is a central depression at least 5 km deep that has a rim showing concentric fabric. The mound surface has a hummocky/blocky surface texture, and is very lightly cratered. A similar but even larger feature (Piccard Mons, PM) is seen in twilight stereo imaging 300 km to the south. This reaches ~6 km high and 225 km across. The general shapes of these edifices and associated structures appear to be constructional. Their origin could involve cryovolcanism (*3*), but entailing materials considerably stronger than $N_2$ ice.

***Tectonics***

Pluto's EH shows numerous belts of aligned, often arcuate, troughs and scarps that can reach several hundred kilometers in length, several kilometers high, and which are often observed to cut across pre-existing landforms as well as branch into one another (Fig. S9, SOM). We interpret these features to be extensional fractures (graben and normal faults) in varying stages of degradation. Notable is the single 3-4-km deep V-shaped trough, Virgil Fossa (VF, Fig. S9C), which has unbroken segments of at least 200 km and an asymmetric upward displacement on the south scarp of 1-2 km. Towards the trough's eastern end it cuts through Elliot crater, and to the west links with a network of smaller, subparallel fractures. The high scarp has an anomalously red color and is associated with water ice (*9*). Other extensional fracture systems are shown in Fig. S9.

Compressional features, if present, are less obvious. One candidate, TD (Figs. S9F and 3B), consists of several elongated swells ~200 km wide, traversed by at least one long extensional feature (Sleipnir Fossa, SF). TD could be due to compressional folding, but may also be analogous to a salt-cored anticline or arch, in which low-density core material contributes to the arching.

The differing fault trends and states of degradation suggest multiple deformation episodes and prolonged tectonic activity. We do not elaborate on their origin here, but note that equatorial normal faults would not arise from despinning stresses (*10*). The great length of individual faults on Pluto, their steepness (>20°, from stereo), spectral evidence (*9*), and the absence of localized flank uplift strongly suggest a thick $H_2O$-ice lithosphere (as opposed to a thin $H_2O$-ice lithosphere, or one made of any of Pluto's volatile ices).

***Impact craters***

Pluto displays a wide variety of crater sizes and morphologies (Figs. S11 and S12; SOM). Globally, recognizable crater diameters range from ~0.5–250 km, not including any possible ancient basin underlying SP. Crater densities vary widely on Pluto, from the heavily cratered portions of CR, to SP, which has no identifiable impact craters. The total cumulative crater size-frequency distribution (SFD) on the EH is shown in Fig. S13A. From this we conclude that Pluto's surface as a whole dates back nearly to the time of the end of Late Heavy Bombardment (LHB), or in the context of the Kuiper belt, the proposed era of rearrangement of the outer solar system (perhaps 4 Gya, e.g., *11*). On the EH, only the eastern portion of CR appears to approach the saturation crater densities (for large craters, cf. Fig. S13B) that would be expected of a terrain that survived from the LHB itself, when cratering rates were likely much higher than at present. In contrast, TD, eastern TR, the water-ice mountain ranges, the mounds (all very lightly cratered), and especially SP (no identified craters) are all very young (Fig. S13C). No craters have been detected in SP down to 2-km diameter, which is a tighter size limit than reported previously (*6*), and implies a model crater retention age of no greater than 10 Ma, and possibly much less (*12*) (SOM).

***Geologic evolution***

Though complex and largely novel, landforms on Pluto present many clues to their origin and history. The basin in which SP is located is ancient, despite the youthfulness of its interior deposits. Its semicircular rim of elevated mountainous terrains suggests that it probably is a heavily modified impact basin. The larger visible craters in these mountainous terrains probably post-date this SP basin.

Except in the west, the uplands surrounding SP have been blanketed with mantles of substantial thickness and various surface compositions (*9*), which have been partially stripped. The primary agents of upland modification probably include sublimation, frost deposition, and glacial erosion. We envision two end-member scenarios: in one, a formerly deep ice mantle (largely $N_2$) covered the uplands surrounding SP but was gradually lost to space. As ice levels dropped, glacial ice eroded the dissected terrains and, to the east of SP, flowed back into SP, leaving remnants in smooth-floored depressions. Alternatively, ices may have been cycled between SP and its surroundings, perhaps episodically, to form the glaciers and dissected terrains. In this case loss to space of volatile ices need not have occurred (*13*). Nitrogen and other volatiles available to the surface environment may also be replenished episodically by sources within Pluto's interior (*14*).

The dark mantles of CR and other local regions conform to present topography, suggesting that they post-date the erosional sculpting of the landscape or are actively recycled. The cellular pattern imposed on SP ices is a relatively young feature, given the absence of craters, and the hectometer-scale pits and ridges on SP comprise the youngest widespread landforms on Pluto.

The relative youth of some extensional features is consistent with predicted recent extensional stresses associated with a late, possibly partial freezing of a subsurface ocean (*15*), though other explanations are also possible. Various lines of evidence, including the spectroscopic identification of water ice along the exposed walls of VF, as well as the steep, chaotic mountains bordering SP, suggest a cold, strong, water ice-based crust.

## Charon

Charon's EH (Fig. 1B) can be divided into two broad provinces separated by a roughly aligned assemblage of ridges and canyons, which span the EH from east to west. North of this tectonic belt is rugged, cratered terrain; south of it are smoother but geologically complex plains. The northern hemisphere is capped by the dark, reddish Mordor Macula (MM). Relief exceeding 20 km is seen in limb profiles and stereo topography (Fig. S16), and is a testament to the bearing strength of cold water ice and Charon's modest surface gravity (0.29 m $s^{-2}$ (*16*)).

### *Cratered northern terrain*

Charon's northern terrain is exceptionally rugged, and contains both a network of polygonal troughs 3-to-6 km deep, and a possibly related irregular depression almost 10 km deep immediately south of the edge of MM near 270°E (Fig. S14). A prominent, ~230-km diameter, 6-km-deep crater **(**Dorothy Gale**)** at 58°N, 38°E (Fig. S2) straddles the discrete edge of MM (*6*). The cumulative crater distribution for Charon's northern terrain is shown in Fig. S14A. The crater density at large sizes, where counts are reliable, implies a surface age older than 4 Gyr (SOM).

The overall dark deposit of MM does not correlate with any specific terrain boundary or geologic unit. A prominent, arcuate ridge ~5 km high ('a' in Fig. S14) coincides with a prominent albedo and color boundary (cf. Fig. S14 with Fig. 1B), and may be an impact basin rim or an extension of the tectonic deformation seen more clearly to the south. Other morphological indicators of an impact origin such as a well-defined ejecta blanket or secondary craters have not been discerned.

***Ridges, troughs, and canyons***

The structural belt that bisects Charon's EH consists of subparallel scarps, ridges, and troughs of variable extent, but over 200 km wide in places (Fig. 1B). Notable are two chasmata: (1) Serenity Chasma, which is >50 km wide and ~5 km deep, and exhibits a pronounced rift-flank uplift; and (2) Mandjet Chasma, which appears to be fault bounded and reaches ~7 km depth (Fig. S2). These chasmata resemble extensional rifts on several mid-sized icy satellites (*10*).

We interpret this assemblage as the structural expression of normal faults and graben that represent substantial, aligned, tectonic extension of Charon's icy crust. Several large craters superposed on the chasmata indicate that this extension is geologically old (see below and Fig. S15). Given the horizontal and vertical scale of these structures, steeply dipping normal faults likely extend to depths of tens of km. They represent global areal extension on the order of ~1%.

***Southern Plains***

The smoother southern half of Charon's EH forms an apparently continuous surface with low relief named Vulcan Planum (VP). Near the bounding scarps to the north, the planum slopes gently downward by ~1 km towards the scarps. Portions of the plains observed at higher resolution exhibit a distinctive, lineated texture of closely spaced grooves or furrows (Fig. 6B). One possible origin for the southern plains is tectonic resurfacing like that seen on the icy satellites Ganymede and Enceladus (*3*). Morphologically distinct groups of deeper, rille-like narrow troughs and furrows that post-date the plains also occur. Although deep, these troughs are nonetheless superimposed by a number of impact craters, and thus are relatively old. The en echelon nature of these troughs, and rough parallelism with the chasmata to the north, suggests a tectonic origin or structural control.

Fields of small hills (2-3 km across), areas of relatively low crater density, and at least one pancake-shaped unit are consistent with cryovolcanic resurfacing (Fig. 6B) (*17*). Peaks surrounded by "moats" (Kubrick and Clarke Montes, KM and CM, see Fig. S2 and 'b' labels in Fig. 6A) were noted by (*6*). The peaks are up to 3-4 km high above the floors of the moats and the moats 1-2 km deep below the surrounding plains. The moat at CM appears to expose a more rugged terrain ('j' label in Fig. 6B), with smooth plains embaying the margins, two of which are lobate. The moats are perhaps due to mountain loading and flexure of Charon's lithosphere. There are two additional depressions surrounded by rounded or lobate margins ('a' labels in Fig. 6A); thus alternatively, both the moats *and* depressions may be the expressions of the flow of, and incomplete enclosure by, viscous, cryovolcanic materials, such as proposed for the uranian moons Ariel and Miranda (*3, 18*).

The SFD of impact craters of the southern plains lies below that for the north at large diameters (≥50 km, Fig. S15A), yet model ages for the plains point to an age of ~4 Gyr (SOM), thus implying an older age for the northern terrain, and a similar or older age for those chasmata that predate (were resurfaced by) VP. In limited regions on VP, however, craters are sparse (Fig.

6B), implying that the resurfacing of VP may have acted over an extended time. The impact SFD of the southern plains is also likely the best representation of the Kuiper belt impact crater production function for the Pluto system (*6*), and one that appears to rule out certain classes of Kuiper belt object population size distributions (SOM).

***Geological evolution***

Charon's surface is dominated by impacts, tectonic deformation, and resurfacing, and as such fits broadly into the accepted picture of geologic evolution on icy satellites (*19*, *20*). That Charon is so geologically complex, however, would seem to require a heat source for reshaping what would have otherwise been a heavily cratered surface. If the ~4 Gyr age of even the youngest of Charon's surfaces is correct, then this activity dates back to an early warmer epoch. The tectonic record is consistent with global expansion, and the smooth plains consistent with the mobilization of volatile ices from the interior. The spatial distribution of tectonic features is not readily reconciled with the kinds of patterns expected (indeed, predicted) from tidal or despinning stresses (*3*). Charon may have had an ancient subsurface ocean that subsequently froze, which would generate the global set of extensional features, and might permit eruption of cryovolcanic magmas (*21*).

## A Divergent Binary

Pluto and Charon are strikingly different in surface appearance, despite their similar densities and presumed bulk compositions (*6*). With the possible exception of MM, the dynamic remolding of landscapes by volatile transport seen on Pluto is not evident on Charon, whose surface is instead dominated spectrally by the signature of water ice (*9*). Whether this is because Charon's near-surface volatile ices have sublimated and have been totally lost to space owing to that body's lower gravity (*22*), or whether something more fundamental related to the origin of the binary and subsequent internal evolution (*23*) is responsible, remains to be determined.

Much of what we see on Pluto can be attributed to surface-atmosphere interactions and the mobilization of volatile ices by insolation. Other geological activity requires or required internal heating. The convection and advection of volatile ices in SP can be powered by present-day radiogenic heat loss (SOM). On the other hand, the prominent mountains at the western margin of SP, and the strange, multi-km-high mound features to the south are both young geologically and presumably composed of relatively strong, water-ice-based geological materials. Their origin, and what drove their formation so late in Solar System history remain uncertain. What is more certain is that all three major Kuiper belt bodies (past or present) visited by spacecraft so far, Pluto, Charon, and Triton, are more different than similar, and bear witness to the potential diversity awaiting the future exploration of their realm.

**Supplementary Materials**

www.sciencemag.org

Materials and Methods

Supplementary Text

Figs. S1 to S15

Table S1

## Acknowledgements

We thank the many engineers who have contributed to the success of the New Horizons mission and NASA's Deep Space Network for a decade of excellent support to New Horizons. We thank reviewer Simon Kattenhorn for his close and meticulous reading, and Pam Engebretson for contribution to figure production. Supporting imagery is available in the supplementary material. As contractually agreed to with NASA, fully calibrated New Horizons Pluto system data will be released via the NASA Planetary Data System at https://pds.nasa.gov/ in a series of stages in 2016 and 2017 as the data set is fully downlinked and calibrated. This work was supported by NASA's New Horizons project.

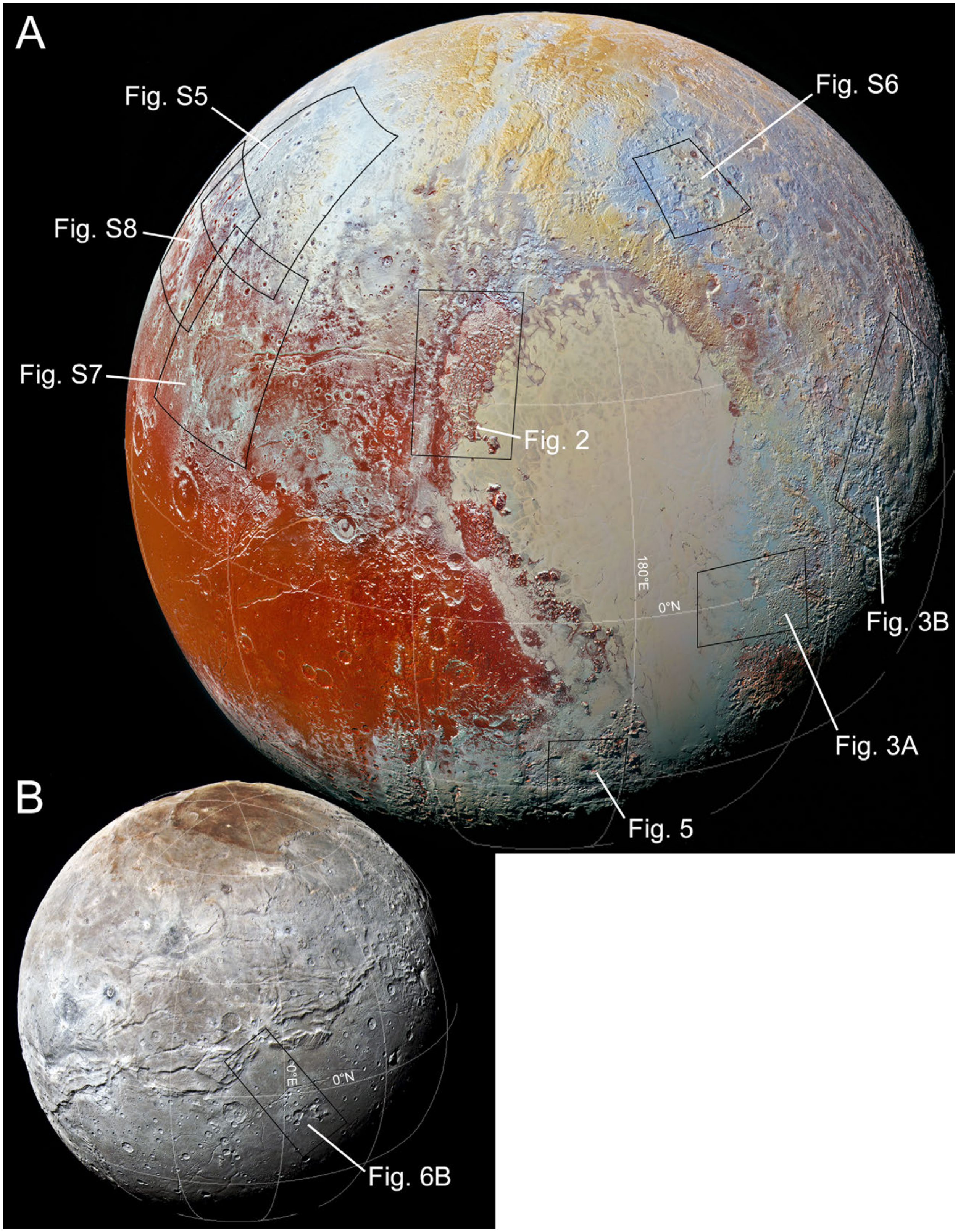

Fig. 1: **Global, enhanced color views of Pluto (A) and Charon (B), with their relative sizes shown to scale.** Filters used are blue-red-near infrared (*5*). Pluto's diameter is 2374 km, and Charon's is 1212 km (*6*). The spacing of the latitude and longitude lines is 30°. Pluto image is 680 m/pixel MVIC coverage of the P_COLOR2 observation, with a sub-spacecraft position of 26.6°N, 167.6°E and a phase angle of 38.0°. Charon image is 1460 m/pixel MVIC coverage of the C_COLOR_2 observation, with a sub-spacecraft position of 25.5°N, 347.5°E and a phase angle of 38.3°. North is up for both. A number of terrains shown in other figures are highlighted and labeled.

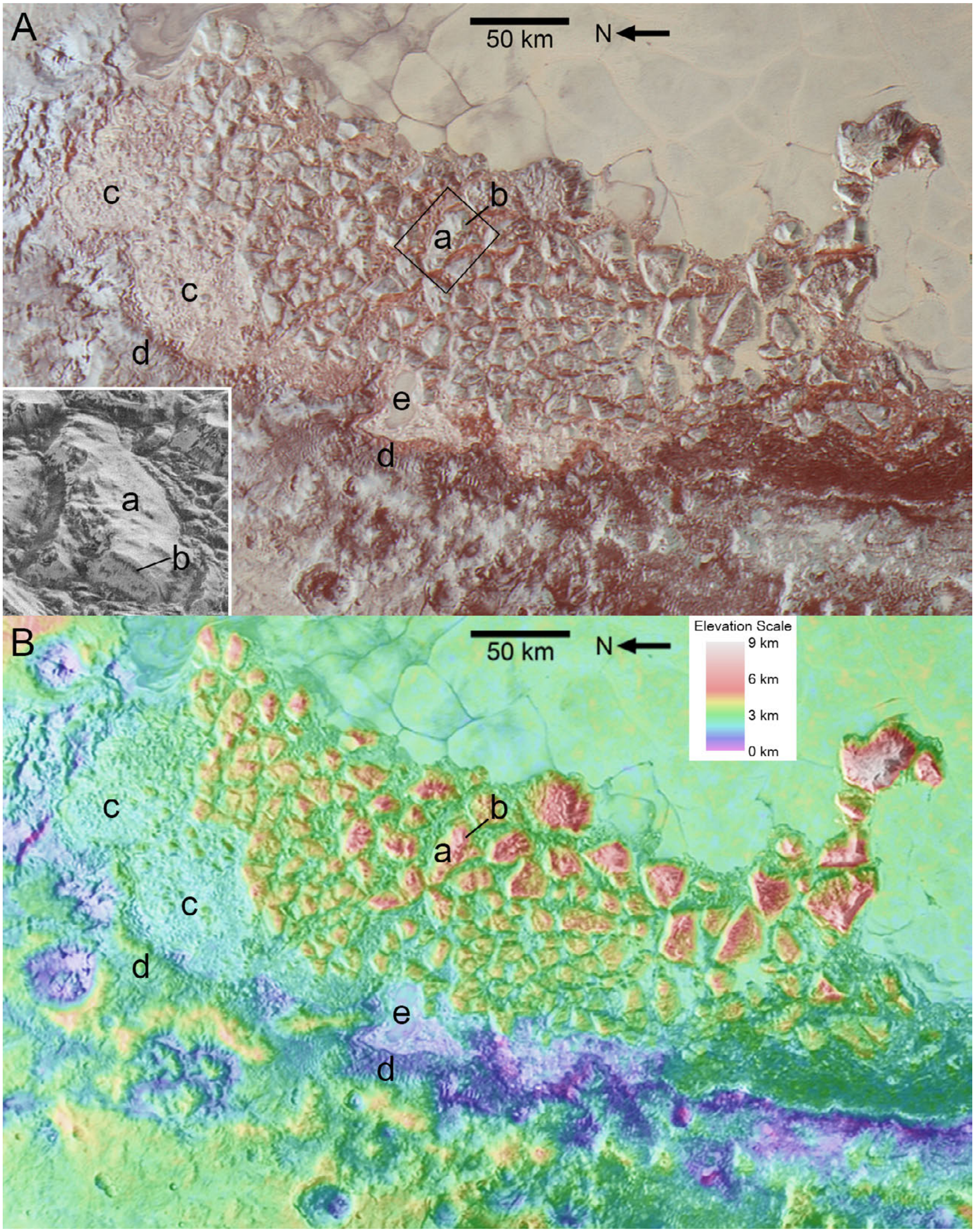

Fig. 2: **The chaotic mountains of al-Idrisi Montes on the northwest margin of Sputnik Planum.** (A) 680 m/pixel, reprojected, MVIC enhanced color coverage of the P_COLOR2

observation, centered at 34.5°N, 155°E. 30 by 40 km-wide inset shows a detail of one mountain in 79 m/pixel, LORRI coverage of the P_MVIC_LORRI_CA observation. (B) Colorized digital elevation model (DEM) overlain on the 680 m/pixel MVIC coverage. (a) Textured surface possibly pre-dating block formation; (b) Steep fracture surface with possible exposed layering; (c) Chaos composed of small blocks; (d) Inward-facing terraces; (e) Small exposure of Sputnik Planum-like material.

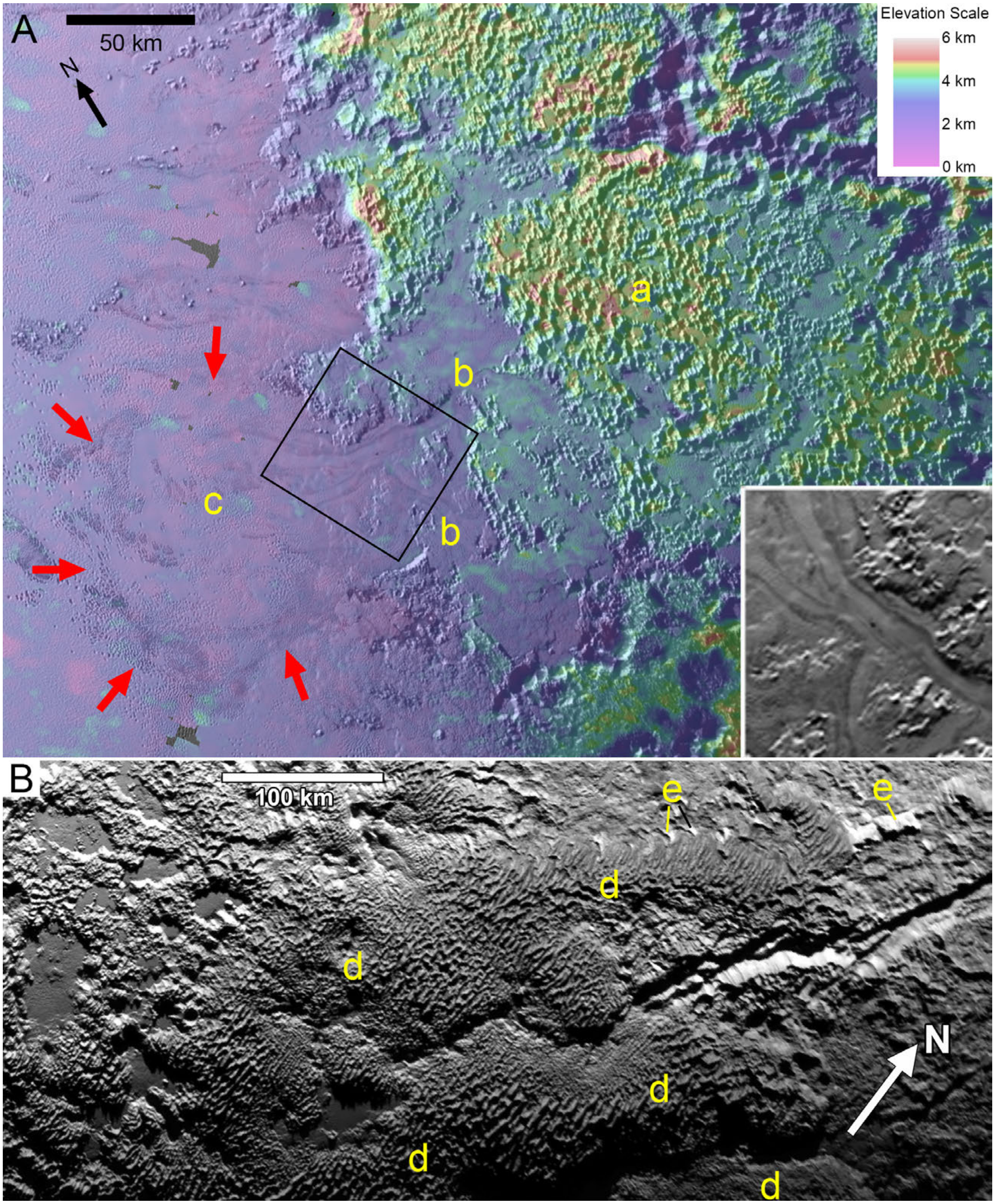

Fig. 3: **Pitted uplands, glaciers, and bladed terrain.** (A) Colorized digital elevation model of pitted uplands and valley glaciers east of Sputnik Planum, overlain on 320 m/pixel, reprojected MVIC coverage of the P_MVIC_LORRI_CA observation, centered at 2°N, 195.5°E. (a) Densely pitted terrain, with smooth material covering the floors of the pits. (b) Smooth plains exhibiting glacial flow through notches in the pitted uplands towards SP. 60 by 50 km-wide rotated inset enhances the contrast of the original MVIC image to emphasize flow lines. (c) Debouchment of a valley glacier into Sputnik Planum, where it assumes the lobate planform of a piedmont glacier. Possible outer flow edges are indicated by red arrows.

(B) Bladed terrain outcropping on top of several broad swells (marked with ‘d’) of Tartarus Dorsa. 680 m/pixel, reprojected MVIC coverage of the P_COLOR2 observation, centered at 17.5°N, 227°E. (e) Triangular and rectangular facets of the plains ramping upwards onto the ridges.

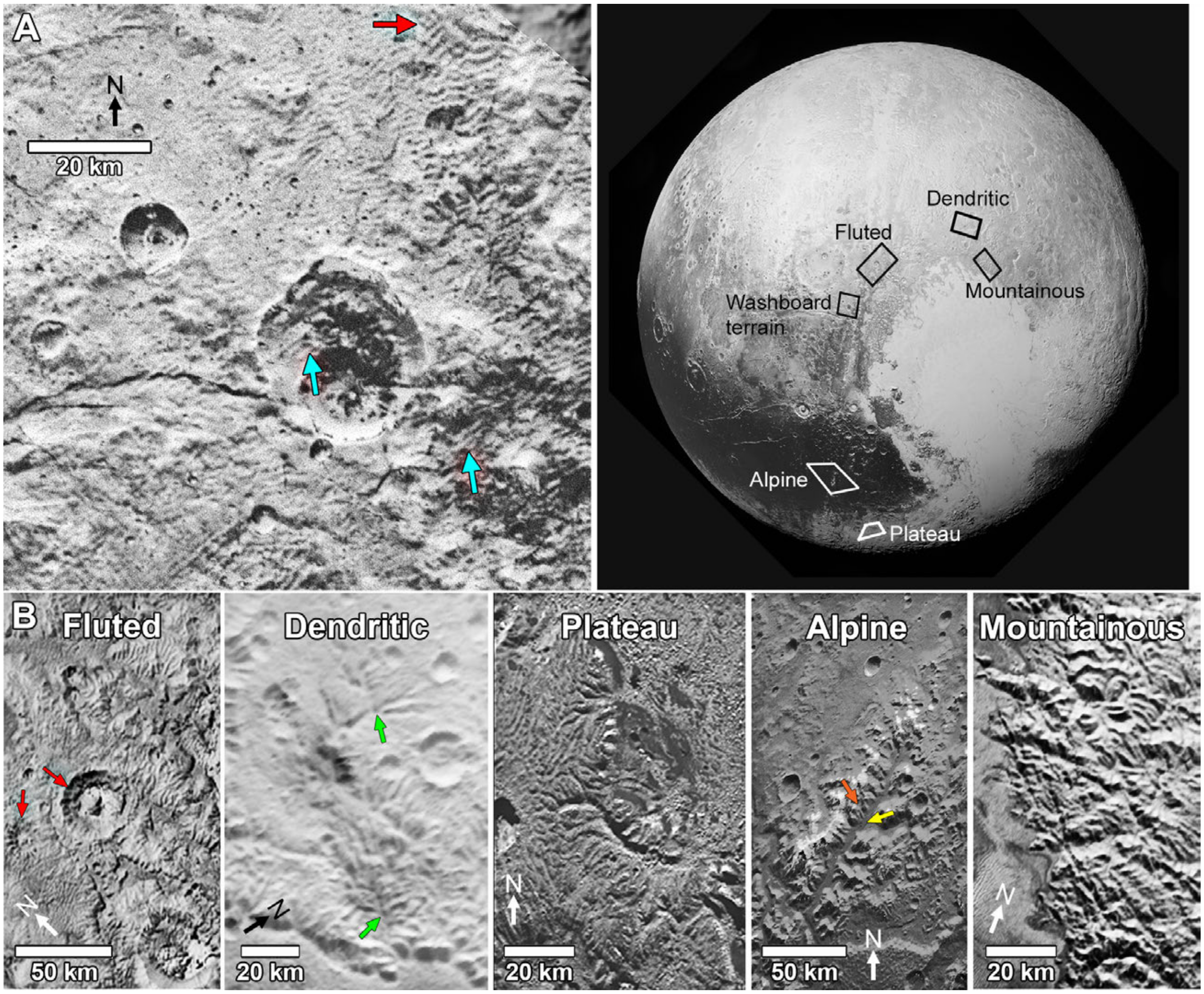

Fig. 4: **Washboard and dissected terrains on Pluto.** Locations of the terrains are highlighted at upper-right. (A) Washboard terrain northwest of Sputnik Planum. 125 m/pixel, reprojected, LORRI coverage from the P_MPAN_1 observation, centered at 38°N, 145.5°E. Blue arrows indicate washboard texture within craters, and the red arrow indicates where washboard terrain has modified fluted terrain.

(B) The five types of dissected terrains with informal typology discussed in the text and SOM. Fluted, dendritic and mountainous terrain images are taken from 680 m/pixel, reprojected MVIC coverage of the P_COLOR2 observation; plateau and alpine terrain images are taken from 320 m/pixel MVIC coverage of the P_MVIC_LORRI_CA observation. The fluted terrain image is centered at 48.4°N, 153.4°E. Red arrows indicate incision of downslope-oriented grooves. The image of dendritic valley networks (green arrows) is centered at 54.8°N, 186.6°E. The dissected plateaus image is centered at 22.1°S, 155.6°E. Alpine valley systems show wide, dendritic trunk valleys (yellow arrow) that head on dissected mountainous slopes (orange arrow); image is centered at 5.2°S, 146.5°E. The mountainous dissection image is centered at 45.4°N, 188.9°E.

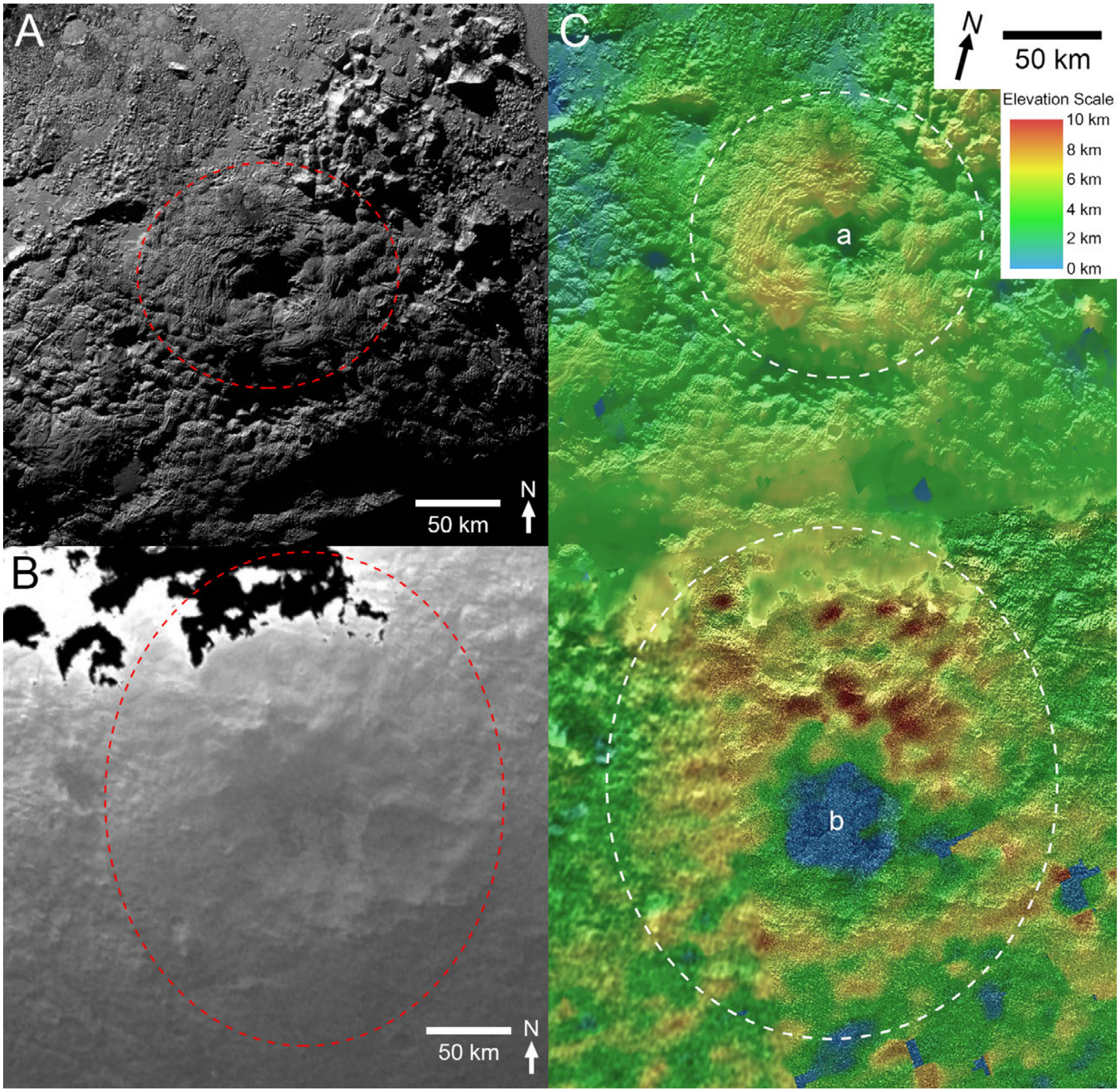

Fig. 5: **Quasi-circular mounds south of Sputnik Planum, both with depressions at their summits, which may have a cryovolcanic contribution.** Dashed lines mark their approximate boundaries. 320 m/pixel, reprojected MVIC coverage of the P_MVIC_LORRI_CA observation. (A) Wright Mons at 22°S, 173°E. (B) Piccard Mons at 35°S, 176°E, seen in twilight. (C) Colorized DEM overlain on the MVIC coverage of the mounds. ‘a’ marks Wright Mons, ‘b’ marks Piccard Mons.

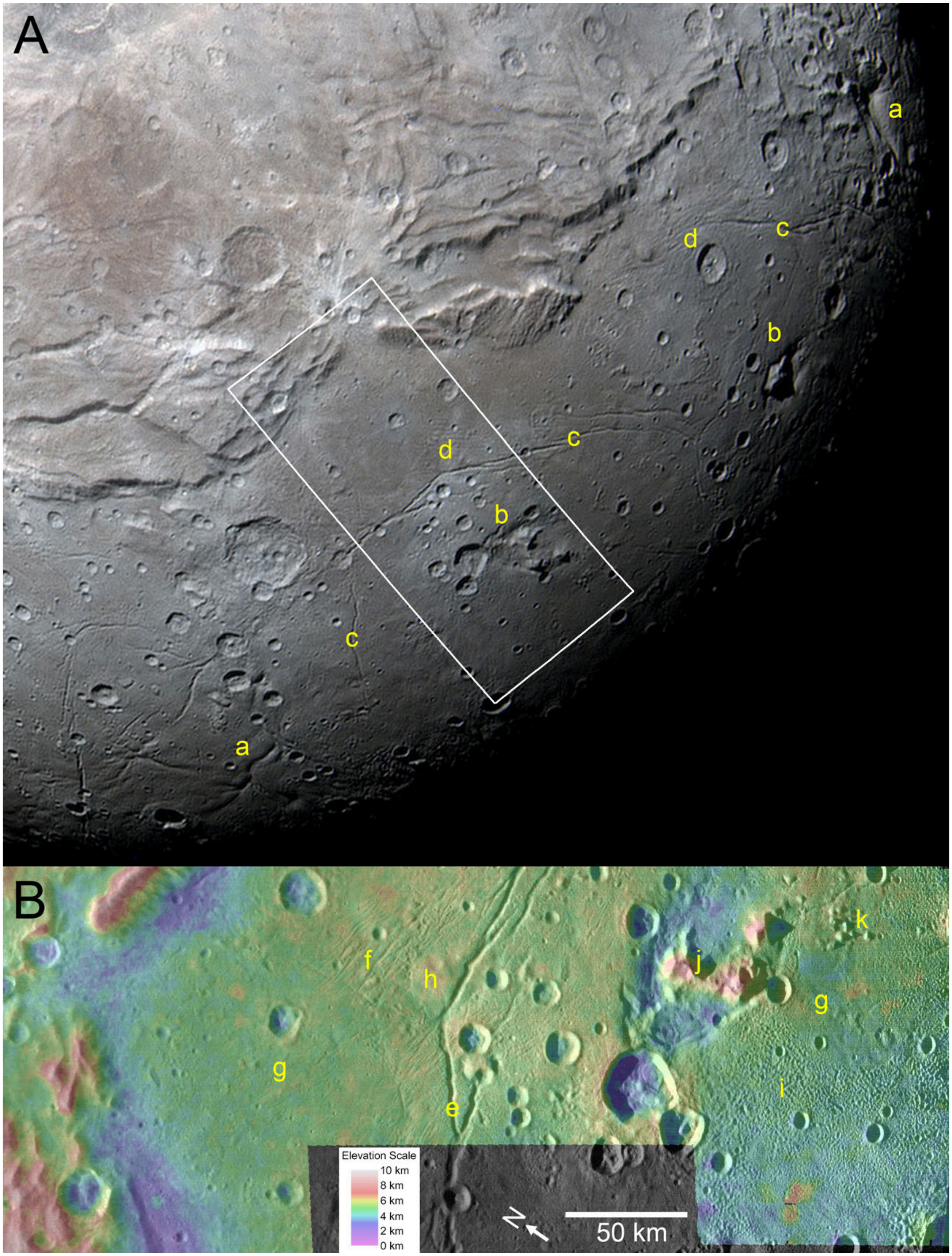

Fig. 6: (A) **Enlargement of a portion of Fig. 1b showing details of Vulcan Planum on Charon.** 1460 m/pixel MVIC coverage is from the C_COLOR_2 observation, centered at 5.5°N, 3°E, with north up. (a) depressions with lobate margins; (b) mountains surrounded by moat-like depressions; (c) deep, rille-like troughs, and (d) shallow, finely spaced furrows. White outline indicates high-resolution image in (B).

(B) **High-resolution view of resurfacing on Vulcan Planum.** 160 m/pixel, reprojected LORRI coverage is from the C_MVIC_LORRI_CA observation centered at 0°N, 0.5°E, with DEM color overlain. Seen in this LORRI view are rille-like troughs (e) and more finely spaced, shallow furrows (f), smoother regions of lower crater density (g), a pancake-shaped deposit (h), and unusual textured terrain (i). Clarke Mons (j) lies in a depression which is itself bordered on two sides by distinctive lobate scarps. (k) A field of small hills.

# Supplementary Materials for

## The Geology of Pluto and Charon Through the Eyes of New Horizons

J. M. Moore*, W. B. McKinnon, J. R. Spencer, A. D. Howard, P. M. Schenk, R. A. Beyer, F. Nimmo, K. N. Singer, O. M. Umurhan, O. L. White, S. A. Stern, K. Ennico, C. B. Olkin, H. A. Weaver, L. A. Young, R. P. Binzel, M. W. Buie, B. J. Buratti, A. F. Cheng, D. P. Cruikshank, W. M. Grundy, I. R. Linscott, H. J. Reitsema, D. C. Reuter, M. R. Showalter, V. J. Bray, C. L. Chavez, C. J. A. Howett, T. R. Lauer, C. M. Lisse, A. H. Parker, S. B. Porter, S. J. Robbins, K. Runyon, T. Stryk, H. B. Throop, C. C. C. Tsang, A. J. Verbiscer, A. M. Zangari, A. L. Chaikin, D. E. Wilhelms

*correspondence to: jeff.moore@nasa.gov

**This PDF file includes:**

Materials and Methods
Supplementary Text
Figs. S1 to S15
Table S1

## Materials and Methods

Acronym List

AIM - al-Idrisi Montes
CM – Clarke Mons
CR - Cthulhu Regio
DEM – Digital Elevation Model
EH - encounter hemisphere
GRAIL - Gravity Recovery and Interior Laboratory
KBO - Kuiper Belt object
KM - Kubrick Mons
LEISA - Linear Etalon Imaging Spectral Array
LHB - Late Heavy Bombardment
LORRI - LOng Range Reconnaissance Imager
MM - Mordor Macula
MVIC - Multispectral Visible Imaging Camera
PM – Piccard Mons
SFD - size-frequency distribution
SF – Sleipnir Fossa
SP - Sputnik Planum
TD - Tartarus Dorsa
TR - Tombaugh Regio
VP – Vulcan Planum
WM – Wright Mons

Derivation of stereo digital elevation models

All of the topography maps (or digital elevation models, DEMs) in this paper have been created using stereo photogrammetry. Our stereo method is an automated photogrammetry package based on scene-recognition algorithms, which attempt to match albedo patterns in finite-sized patches in each of two stereo images that observe a particular area on a planetary surface at different angles, from which parallax and the corresponding difference in elevation between points can be determined. This technique has been successfully applied to the Galilean and saturnian satellites (e.g. *24-29*), and we refer the reader to these references for detailed information on its construction and implementation. Stereo DEMs are reliable at regional scales, but are unable to resolve fine scale topographic features that are smaller than approximately five times the resolution of the poorest resolution image in the stereo pair due to pixel averaging. The technique is most effective for terrains that display pronounced albedo contrasts across a small spatial scale; low-contrast, featureless terrain will confuse the algorithm's attempt to match albedo patterns in the two stereo images. On Pluto, this becomes an issue when the technique is applied to low-contrast terrains including the high-albedo center of Sputnik Planum, and the low-albedo Cthulhu Regio, but outside these areas high quality DEMs are obtained.

We present data from three separate DEMs across six figures in this paper and the SOM: Figs. 2B, 3A, 5C, 6B, S6, and S14. Individual observations during the encounter are designated by name (e.g. P_MVIC_LORRI_CA):

- The DEM used in Figs. 2B, 3A, 5C and S6 was created using the MVIC P_MPan1 (495 m/pixel resolution) and MVIC P_MVIC_LORRI_CA (320 m/pixel resolution) observations as left and right images in the stereo pair. The resulting DEM can resolve features that are at least 2.5 km across, and has a vertical precision of 225 m.
- The DEM used in Fig. 6B was created using the MVIC C_MVIC_LORRI_CA (628 m/pixel resolution) and the LORRI C_MVIC_LORRI_CA (157 m/pixel resolution) observations as left and right images in the stereo pair. The resulting DEM can resolve features that are at least 3.1 km across, and has a vertical precision of 110 m.
- The DEM used in Fig. S14 was created using the MVIC C_Color_2 (1463 m/pixel resolution) and the LORRI C_LORRI (888 m/pixel resolution) observations as left and right images in the stereo pair. The resulting DEM can resolve features that are at least 7.3 km across, and has a vertical precision of 1250 m.

**Supplementary Text**

Sputnik Planum

The high albedo of the plains of Sputnik Planum (all names used on Pluto and Charon are informal) is consistent with a composition dominated by $N_2$ ice. $N_2$, CO, and $CH_4$ have all been detected within Sputnik Planum by the LEISA (Linear Etalon Imaging Spectral Array) spectral imager, although the relative amounts have not yet been modeled (*9*). Globally, however, $N_2$ is the dominant surface ice on Pluto (*30*), and we take nitrogen-dominated ice within Sputnik Planum as our working hypothesis.

The central and northern regions of Sputnik Planum display a distinct cellular morphology, which varies in appearance across the Planum. In the bright central portion of the Planum (Fig. S4A, corresponding to a region relatively high in CO ice as indicated by LEISA (*6*, *9*)), the cells are bounded by troughs that typically reach a few kilometers across and around a hundred meters deep, and which often display central ridges reaching ~1 km across. In some instances, dark material collects within these troughs. The cellular morphology is most apparent at the southern margin of this region (Fig. S4A). In the center of Sputnik Planum, the trough network becomes less interconnected. In the north of Sputnik Planum, where the cell interiors appear darker (Fig. S4B), cell boundaries are represented less by sharply-defined troughs and more by bright, diffuse lineations, which show no apparent relief, possibly due to infilling or relaxation. At the northernmost margin of Sputnik Planum (Fig. S4C), still darker ice is encountered, where it concentrates around the cell boundaries, which here appear thin (~1 km wide) and dark and show no apparent relief. Indications of flow in the form of swirl patterns and topographic embayment are apparent here (*6*), as well as near Hillary Montes (~5° N, 170° E) and between these Montes and Cthulhu Regio to the west (Fig. S1C).

Ice plains within the southernmost region of Sputnik Planum exist as two large lobes extending either side of Norgay Montes (~15° S, 175° E, Fig. S1C). The plains here, and on the eastern and western margins of Sputnik Planum, do not display cellular morphology, instead appearing featureless or showing dense concentrations of pits (Fig. S4D). Individual pits reach a few kilometers across, and show planforms ranging from equidimensional to elongate. Sublimation of the plains ice may form these pits, perhaps through a process similar to formation of the 'Swiss cheese features' in the south polar $CO_2$ ice cap of Mars (*31*); nitrogen ice on Pluto plays an analogous role to $CO_2$ ice on Mars, maintaining Pluto's atmosphere in vapor pressure equilibrium. Pits in Sputnik Planum are seldom seen within the cellular terrain relative to the southern and eastern plains, perhaps indicating that convection within the cellular terrain allows resurfacing of the ice on a timescale sufficient to eliminate any pits that form due to sublimation; the dense concentrations of pits within the southern and eastern plains would therefore imply that these plains are comparatively older than the cellular terrain.

These plains also feature streaks that appear to be composed of dense clusters of aligned pits with dark material covering their floors, and which seem to define the outlines of elongated cells, but which do not display any apparent relief (left-hand side of Fig. S4D). These 'ghost cells' are adjacent to the cellular terrain of Sputnik Planum, and the dark material may once have filled the troughs of actively convecting cells, which have since relaxed and lost their relief. Because of the noticeably stretched planforms of the pits and cell outlines in a ~N-S direction, the loss of relief may have been due to the nitrogen ice having flowed southward, thereby moving away from heat flow or other conditions that permitted convection within the cellular terrain of Sputnik Planum. Without convection to renew the surface and maintain cell relief, the dark material that once filled the troughs remains as the only indicator of the cells' existence, and its low albedo may cause heating of the nitrogen ice below, promoting sublimation and pit formation. The lack of a deeper heat source would, in contrast, help to maintain the relief of these pits by inhibiting their closure via inwards flow of bright, cold nitrogen ice.

In the eastern portion of Sputnik Planum where it borders the pitted uplands, numerous isolated hills that typically reach a few km across are scattered across the plains (Fig. S4E). They often form densely packed, generally elongate clusters that reach up to 20 km long, and, in the case of Challenger Colles (see Fig. S1B), form a rectangular, well-defined unit that measures 60 km by 35 km. Where these hills occur in the cellular terrain, they are almost exclusively coincident with the cell boundaries. Virtually no hills are seen in Sputnik Planum beyond ~250 km from the boundary with the pitted uplands. Close to the pitted uplands, groups of these hills tend to form chains that are coincident with the paths of glaciers emanating from the uplands. Isolated hills within smooth plains seen in the pitted uplands are hypothesized to be embedded fragments of the pitted uplands material (see main text section on this subject), and in Sputnik Planum the chains of hills may represent such fragments that have been carried out of the uplands by the putative glaciers into Sputnik Planum. If so, they may well have a water ice-based composition, allowing them to float on the nitrogen and/or carbon monoxide ice (which have very similar densities), as the chaotic mountains and smaller blocks on the western margin of Sputnik Planum are hypothesized to do (see next section). Once the fragments transfer from the glacial regime to the cellular regime, they would be subject to the solid-state convection motions of the nitrogen ice, and would move to the cell boundaries and congregate into densely packed groups. Challenger Colles may be an especially large accumulation of these fragments, perhaps where many of them have become grounded on a shallow substrate, one beneath the nitrogen ice-dominated layer.

Isostasy in Sputnik Planum

Water ice Ih has a solid density of 0.935 g/cm$^3$ in the temperature interval between 40 and 100 K (*32*). Water ice is elastic and brittle at those temperatures, so fractures and porosity created through geological processes (such as the prominent impact cratering on the west margin of SP or tectonics) can be maintained through geologic time as long as pressures and temperatures remain modest (*33*). We note for comparison that GRAIL (lunar mission) gravity data has conclusively shown that the lunar crust sustains porosities of 4–20% to considerable depths (at least several kilometers) (*34*), consistent with its bombardment history. Pluto's upper water ice crust, as exposed in its prominent mountains, could thus be underdense compared with solid ice, with densities between 0.90 and 0.75 g/cm$^3$ for a similar range of crustal porosities as on the Moon.

Solid $N_2$ densities vary between 1.00 g/cm$^3$ at 36 K (close to Pluto's present nitrogen ice surface temperature) and 0.942 g/cm$^3$ at the $N_2$ melting temperature at vapor pressure equilibrium of 63.15 K (*37*). Nitrogen ice is deformable enough at Pluto surface conditions (*7, 8*) that below any near-surface frost layer it will anneal into a coherent, crystalline solid. Hence whether solid or porous, water ice blocks will be buoyant in a layer of solid $N_2$.

For water ice blocks of 0.80 g/cm$^3$ (15% porosity) floating in isostatic equilibrium in "warm" but solid $N_2$ of 0.96 g/cm$^3$, emergent topography could be supported by ice keels of 0.8/(0.96 – 0.8) = 5 times vertical extent, if the topography were mirrored. Of course, the actual keel necessary for isostasy would depend on the specific keel topography, but this calculation suggests that ice mountains on Pluto up to a kilometer in height could easily float, or have floated, on a layer of solid $N_2$ a few km deep. Indeed, the mountains within Sputnik Planum rather resemble the angular blocks in europan chaos, i.e. fragments of pre-existing ice crust that have been detached by fracturing, transported, and rotated (*35*).

The origin of the basin in which Sputnik Planum is located is unknown, but it may be the site of an ancient impact basin (*36*), in which case a depth of a few km is quite plausible (for comparison, the observed depth of the largest confirmed impact crater on Charon is 6 km; see main text). On the other hand, the requirements for isostasy of the highest peaks found within Sputnik Planum (those in excess of 3-km height) are more severe (keels possibly in excess of 15 km), and likely exceed any plausible depth for the volatile ices that make up Sputnik Planum. There is, however, the possibility that relatively dense volatile ices accumulating within a broad impact basin could lead to isostatic adjustment of Pluto's crust as a whole, in the manner of accumulating sediments in the Gulf of Mexico and sedimentary basins elsewhere on Earth. In this case the total thickness of Sputnik Planum's volatile ices could be much larger than estimated from pristine basin depths on other icy bodies (*26*). An important caveat, which might limit deep $N_2$ ice layers, is that in the absence of solid state convection (discussed below), $N_2$ ice should melt at depths of a couple of kilometers or so for present-day heat flows (*38*), and liquid $N_2$ is relatively buoyant (0.8 g/cm$^3$).

*Other volatile ices in Sputnik Planum*

As noted above, CO and $CH_4$ ice have also been identified within Sputnik Planum. Spectroscopy alone cannot, however, determine the bulk composition of SP's volatile ices at kilometers depth. Nitrogen is the most volatile of these ices, is the major constituent of Pluto's sublimation-supported atmosphere (>99% (*13*)), and therefore must be the principal material component in Pluto's glacial cycle. Arguments developed in this paper for the buoyancy of water-ice-rich blocks and the piedmont-glacier-style flow of volatile ice onto the surface of SP (Fig. 3A), as opposed to the latter's sinking into the Planum as a density current, imply that the Planum is not dominated by low-density $CH_4$-rich ice (density ≈ 0.5 g/cm$^3$). Hence in discussing SP we focus on $N_2$-dominated ices, while recognizing that CO-rich ice may be important at depth in the SP volatile-ice layer. CO and $N_2$ share the same molecular weight, and have numerous solid-state properties in common (*39*). As crystalline solids they form a nearly complete solid solution with each other (*40*) and the density of CO ice is only about 1.5% greater than that of $N_2$ ice over the temperature range of interest (*39*). So in terms of buoyancy or bulk thermal properties, we can treat $N_2$ and CO as the same.

## Solid-state Convection

The cellular pattern observed over much of Sputnik Planum is suggestive of convective upwellings and downwellings (*6*), and we show here that the rheology of weak, van der Waals bonded ices such as $N_2$ (or CO) likely permits solid-state overturn within a multi-km thick layer on Pluto.

$N_2$ ice in laboratory experiments exhibits uniaxial strain rates at 56 K (0.89 $T_m$, where $T_m$ is melting temperature) of 2 x $10^{-4}$ s$^{-1}$ at differential stresses of 100 kPa (1 bar) (*8*), equivalent to a viscosity $\eta$ ~2 x $10^8$ Pa-s. This is orders of magnitude less viscous than water ice near its melting temperature at similar stresses (~$10^{13}$ Pa-s for temperate glacier ice (*41*)). The viscosity of $N_2$ ice is also temperature dependent. For the activation energy $Q*$ assumed for volume diffusion of 8.6 kJ/mole (*7*), the natural exponential viscosity temperature scale $\delta T = RT^2_{ad}/Q*$, where $R$ = the gas constant and $T_{ad}$ is the adiabatic temperature of the potentially convecting layer, is only 3 K if $T_{ad}$ = 56 K (see, e.g., *42-44*). The temperature difference across the layer, $\Delta T$, which would drive the convection, is limited by the long-term average surface ice temperature (perhaps 36 K (*45*)) at the top and the $N_2$-ice melting temperature (63 K) at the bottom, or ≈ 27 K. This maximum $\Delta T$ in turn implies a maximum (linearized) viscosity ratio ($\Delta\eta$) from top to bottom of $\exp(\Delta T/\delta T) \sim 10^4$.

Activation energies were not derived in (*8*), but their data indicate a smaller $Q^*$ (see next section) and thus a larger $\delta T$ (~6K) and smaller $\Delta\eta$ than suggested above. In either case, the viscosity ratio across a layer of solid nitrogen will be limited and not be high enough to place it into the stagnant lid convective regime (*43, 44*). Most importantly, for a range of Rayleigh numbers *Ra* (the dimensionless measure of the vigor of convection) above the critical value for the onset of convection, $Ra_{cr}$, convection proceeds in the so-called sluggish or mobile lid regime. In this regime, the surface is in motion and transports heat, but moves at a much slower pace than the deeper, warmer interior. As such it is characterized by large aspect ratio convective cells (e.g., *46*), which would be consistent with observed cell dimensions of 20-30 km if the Sputnik Planum layer thickness were at least several kilometers.

The rheologies measured by (*8*) are mildly non-Newtonian (stress dependent with power law indices $n$ ~2), and as such it is important to consider convective stress levels ($\sigma$). These are given by $\sigma \sim 10^{-1} \times \alpha\rho g h \delta T$, where $\alpha$ is the volume thermal expansion coefficient ($2 \times 10^{-3}$ $K^{-1}$ for $N_2$ ice (*37*)), $g$ is surface gravity (0.617 $m/s^2$), and $h$ is the vertical scale associated with $\delta T$, often the convective boundary layer thickness, i.e., a length scale smaller than the entire layer depth (*47*). Table S1 shows estimates of $\sigma$ for some plausible combinations of $h$ and $\delta T$, and from these, $\eta$ values are derived from the flow laws measured for $N_2$ ice at either 56 K or 45 K (*8*). Such viscosity values in turn allow estimates of basal $Ra_b$, using these $\eta$ and, for concreteness, a layer depth $H$ of 4 km (note that $Ra_b$ increases as $H^3$). The $Ra_b$ values in Table S1 are all well above the $Ra_{cr}$ for sluggish lid convection (*44*). Moreover, an empirical scaling for sluggish lid convection (*48*) can be used to estimate the surface heat flow $Nu \times k\Delta T/H$, where $Nu$ is the Nusselt number, the dimensionless measure of heat flow, and $k$ is the thermal conductivity (0.2 W/m·K for $N_2$ ice (*37*)). For example, assuming $Ra_b = 10^6$ and $\Delta\eta = 10^3$, scaling gives $Nu \sim 4.5$, which translates to a heat flow of ~ 4.5 mW/m$^2$ for a total $\Delta T$ of 20 K over 4 km. This is close to the expected radiogenic heat flow from Pluto today, several mW/m$^2$ (*15, 49, 50*).

Comparing this latter value with the heat flow estimates in Table S1, we see some estimates exceed Pluto's likely present-day heat flow. This can be interpreted in two ways. One is that the $T_b$ assumed in a given case in Table S1 is too high. But another is that solid state convection in $N_2$ ice occurs readily, and may occur in layers much thinner than 4 km, and thus would be likely capable of transporting even higher heat flows, such as those in the distant geologic past.

The values of $Ra_b$ and $\Delta\eta$ in Table S1 are not meant to precisely define the convective parameters for Sputnik Planum. Rheological understanding of $N_2$ ice is simply too limited. For example, volume diffusion has been previously argued on theoretical grounds to be the dominant creep mechanism in $N_2$ ice (e.g., *38*), although unless the ice grain sizes are very small, it is more likely that the relatively soft, power-law creep actually measured by (*8*) will be the dominant mechanism. In addition, CO ice readily forms a solid solution with $N_2$ ice. A sufficient concentration of CO would imply an orientationally ordered, $\alpha$ crystalline phase at Pluto temperatures, as opposed to the disordered, $\beta$-phase characteristic of $N_2$ ice at the same temperatures (*39*). The rheological properties of $\alpha$-phase CO ice have not been measured to our knowledge, but based on the rheologies of ordered and disordered water ice phases (*51*), we anticipate that it is more viscous than $\beta$-$N_2$ ice at the same conditions.

Thus, the examples in Table S1 are simply meant to illustrate the plausibility of convective heat transport today within a sufficiently thick layer of relatively deformable, volatile ice. A more rigorous evaluation of convective modes on Pluto must also explicitly take into account the non-Newtonian character of $N_2$ ice creep. The same division into isoviscous, sluggish lid, and stagnant lid convective regimes occurs for non-Newtonian rheologies (*52*).

Glacial Flow of Nitrogen Ice

Owing to its weak crystalline bonding (noted above), $N_2$-dominated ices are expected to flow relatively quickly on the surface of Pluto. Much of the high relief topography observed on Pluto's surface is not expected to be supported by such material (*7*). In the following we examine the relaxation and flow timescales of $N_2$ ice based on available laboratory data.

As noted above, the steady-state creep of $N_2$ is non-Newtonian. The measured stress ($\sigma$) and strain-rate ($\dot{\epsilon}$) obey the power-law relationship (*8*):

$$\dot{\epsilon} = A(T)\sigma^{n(T)}, \qquad \text{(S1)}$$

with a temperature dependent pre-factor $A$ and exponent $n$. The tangential, shear stress $\tau$ due to surface relief on Pluto is a function of the overlying mass and angle of the sloping surface. Assuming the pre-factor is an Arrhenius type of function of the form:

$$A(T) = A_{45} \exp\left(\frac{T_A}{T} - \frac{T_A}{45\ \mathrm{K}}\right) \qquad \text{(S2)}$$

where $A_{45} = 0.005\, s^{-1}(\mathrm{MPa})^{-n(T)}$, fitting the data reported in Table 1 of (*8*) shows that the corresponding approximate value of the activation temperature is $T_A$ = 422 K (equivalent to a $Q^*$ = 3.5 kJ/mole). The value of the exponent $n$, which hovers around 2.1 in the temperature range of interest, suggests that the underlying creep mechanism may be grain boundary sliding (e.g. *53*). A simple linear fit of the same data set gives $n(T) = 2.1 + 0.0155\,(T/\mathrm{K} - 45)$.

The inverse of the strain rate, $\dot{\epsilon}^{-1}$, gives a measure of the deformation timescale of ice structures. Assuming an ice layer of thickness $H$ that is set at an angle $\theta$ with respect to the horizontal, the shear stress of the ice at the base of its layer is $\tau = \rho g H \sin\theta$. Taking the density of $N_2$ ice to be 986.5 kg/m$^3$ (its value at $T$=45K (*37*)), we can estimate the strain or deformation rate for $N_2$ ice structures by combining Eq. (1.1) with $\tau$ given together with the fitted form given in Eq. (1.2) to find:

$$\dot{\epsilon} \approx 442\mathrm{yr}^{-1} \cdot \left(\frac{H}{100\mathrm{m}}\right)^{n} \cdot (sin\theta)^{n} \cdot \exp\left[\frac{T_A}{45\ \mathrm{K}} - \frac{T_A}{T} - 0.091\mathrm{K}^{-1}(T - 45\ \mathrm{K})\right] \qquad \text{(S3)}$$

Thin ice layers with low sloping angles may be treated as a “shallow” flow since the magnitude of horizontal velocities $u$ dominates vertical velocities $w$ under such conditions (*54*, *55*). Further, assuming a 1D flow with uniform interior temperature means one may write $\dot{\epsilon} = du/dz$ and vertically integrate Equation (S1) once to get an estimate for the horizontal flow speeds on the surface, which is given by $u_H \approx \dot{\epsilon} H/(n+1)$.

The above forms for both $\dot{\epsilon}$ and $u_H$ allows us to make certain estimates. For example, we consider the time it takes a sloping channel with a given uniform depth of ice to completely drain itself out assuming this process takes place in the absence of basal melt. For $N_2$ ice at T=45 K and at a depth of 100 meters, together with a sloping angle of $\theta = 10°$, yields a unit strain time of $\dot{\epsilon}^{-1} \approx 1/11.5\ \mathrm{yr}$, or slightly longer than an Earth month. The surface of such an ice mass flows at an approximate rate of $u_H = 375\ \mathrm{m/yr}$, suggesting that the drainage time of a 10 km long channel at this grade is about 26 yr. By contrast, an ice layer with these dimensions but with a lower temperature $T$=38 K instead relaxes on a time scale of $\dot{\epsilon}^{-1} \approx 0.21\ \mathrm{yr}$, with a surface flow speed of $u_H = 154$ m/yr. A similar ice layer, but with a much shallower slope angle $\theta$ = 0.5°, gives corresponding deformation timescales and flow rates of 84 yr and 0.39 m/yr respectively, whereupon a 100 km long basin is traversed by the surface flow in about 250,000 yrs. Inferred rates are very sensitive to the flow thickness assumed – it goes as roughly $H^3$ in this case.

The timescales quoted above are likely upper limits because the strongly insulating nature of $N_2$ ice will imply warmer interior temperatures. Temperature gradients through pure $N_2$ ice could be 20-25 K/km, given estimates of Pluto's current geothermal heat flux of ~4 mW/m$^2$ (*15, 49, 50*). The time it takes to achieve such a state can be estimated by the time it takes a simple thermal wave to traverse up a layer of thickness H with thermal diffusivity $\kappa = K/\rho C_p$. This gives an estimate for the thermal timescale of $t_{\rm th} = H^2/\kappa$. For a heat capacity $C_p \approx 1500$ J/kg·K (*37*), the thermal equilibration time for plutonian $N_2$ ice is $t_{\rm th} \approx (H/2m)^2$ yr.

This underscores the importance of assessing the timescales on which $N_2$ is deposited on the surface. If the surface deposition occurs at a rate of 0.5-2 cm/yr (*9*), then according to the above, ice layers grown at this rate will always be in thermal equilibrium given the emergent interior geothermal heat flux. A given vertically thick ice layer grown in this way will move faster than if the same layer had been placed in-situ all at once because the larger interior temperatures at the base in the slowly grown ice layer scenario imply much lower basal viscosities and, hence, higher flow rates.

Formation of Washboard Terrain

The regular, parallel patterning of the washboard terrain (Fig. 4A), with inter-ridge spacing of about 1 km and ridge lengths often of tens of kilometers, is difficult to explain. Sublimation can produce regularly-spaced landforms like penitentes and snow cups, but they tend to be nearly equidimensional (*56-60*). The large scale of the washboard texture is also problematic when compared to the meter to decimeter scale of terrestrial ablation topography. However, in many geomorphic systems large features can dominate and cannibalize initially smaller features if sufficient time and space exists. The lateral continuity of washboard ridges suggests if sublimation were the dominant process, the washboard crests would have had to migrate over long distances. Such long-distance migration through sublimation has produced the parallel troughs in the martian polar caps (*61*, *62*). We see no evidence, however, for systematic lateral migration in the case of the washboard terrain. If the sublimation process were influenced by persistent atmospheric circulation, lateral continuity could be related to wave-like structures and the interaction of sublimation and wind. The rarified modern atmosphere of Pluto, however, is unlikely to host such structures, but a denser prior atmosphere might also permit ice particle transport and transverse dune formation (*3*).

Alternately, the washboard terrain could potentially be related to the former presence of an ice cover over the landscape. An unusual type of low relief ground moraine topography has formed beneath some ice sheets on Earth called washboard or corrugated moraine (e.g., *63*), which is characterized by parallel ridge and swale topography (albeit with a shorter wavelength of 30-200 m). Its mode of origin is uncertain, but (*63*) suggest it forms beneath surging glaciers, due to formation of crevasses and injection of sediment from below. The origin of the consistent NE-SW orientation of ridges in Pluto's washboard terrain is also uncertain, but could be related to solar illumination, atmospheric circulation patterns, or glacial flow directions.

Formation of Dissected Terrain

As described in the main text, we identify five categories of dissected terrains on Pluto, examples of which are shown in Fig. 4B. Already discussed in the main text are fluted terrain, characterized by incision of downslope-oriented grooves at ~3 km spacing (red arrows), and dendritic valley networks (green arrows), which have branching linear depressions with deeper incision along the trunk valleys, somewhat analogous to terrestrial fluvial drainage basins. The three remaining varieties include dissected plateaus, which also have dendritic structure, but are characterized by modestly incised valleys on upland plateaus with a palmate structure emptying into deeply-incised trunk valleys; alpine valley systems, which are characterized by wide, dendritic trunk valleys (yellow arrow) that are straight to gently curving and which head on dissected mountainous slopes (orange arrow), that have a style of dissection similar to that of the fluted terrain; and mountainous dissection, which is characterized by steep, branched valley networks giving the appearance of dissected terrestrial mountain chains.

A variety of processes form valley-like morphology on Earth and the other planets. We consider below the major categories of such processes and evaluate whether they are possible contributors to one or more of the types of dissected terrains on Pluto.

1. Shear failure of the underlying materials. This could occur either through slow flow, as in terrestrial earthflows, which occasionally form dendritic complexes (*64*), or through failure due to rapid avalanching, such as occurs on terrestrial mountain slopes (*65*) and spur-and-gully terrain on the walls of the martian Valles Marineris. Triggering of such flows generally requires a critical slope gradient. Plutoquakes could also provide critical stresses.

2. Accumulation and avalanching failure of ices accumulated from the atmosphere. Snow avalanches contribute to chute erosion on terrestrial mountain slopes (*66*), and avalanching of seasonal $CO_2$ snows is one suggested mechanism for formation of martian gullies (e.g., *67*).

3. Surface ice accumulation, glacial flow, and erosion. This is the dominant mechanism for erosion by valley glaciers on Earth (e.g., *68, 69*).

4. Erosion beneath thick ice sheets, such as ice streams at the margins of terrestrial plateau glaciers, such as at Greenland and Antarctica (e.g., *70*).

5. Erosion by precipitation of volatiles such as rain or snow, followed by erosion by liquid runoff, which accounts for most terrestrial valley networks (e.g., *71*).

Mechanisms 1 and 2 could potentially contribute to formation of the steep fluted terrain, but are unlikely to contribute to erosion of the larger dendritic valleys. In addition, these mechanisms tend to deposit transported material close to their sources as fans or lobate deposits. Although image resolution may limit recognition, such depositional forms have not been identified. Terrestrial valley glaciers are competent agents of erosion, long-distance transport, and deposition. Most terrestrial valley glaciers occupy former fluvially-sculpted valley networks (e.g., *72*); whether ice accumulating on undissected uplands would sculpt dendritic valley networks is uncertain. Most terrestrial valley glaciers are warm-based, with meltwater production that contributes to basal sliding and accompanying substrate erosion by abrasion and quarrying (e.g., *73, 74*). Cold-based ice sheets on Earth and Mars have generally been considered to be inefficient in eroding their beds (*75, 76*). Recent studies, however, suggest that cold glacier beds can be eroded at modest rates by plucking and bed shear (*77, 78*). Erosion by nitrogen ice flowing over a water ice substrate on Pluto might be aided by the lower density of the latter. Thick enough accumulations of nitrogen-rich ices on Pluto (see Glacial Flow of Nitrogen Ice section) can result in basal melting under reasonable thermal gradients, which can lubricate the base of the glacier and substantially increase its capacity for erosion. Stereo imaging in the dendritic valley network and mountainous terrains suggests some valleys end in depressions or have irregular profiles, which is a characteristic of terrestrial valley glacier erosion (although post-formation processes may have affected the Pluto valleys). The valley networks of the plateau-dissected terrain on Pluto are suggestive of some terrestrial glacial valley networks with broad, deeply incised trunk valleys and hanging tributaries.

Erosion by precipitation and runoff is impossible in Pluto's current environment, and would require a dramatically thicker past atmosphere. Thus we conclude that the most reasonable explanation for formation of Pluto's dissected terrains is glacial erosion by former accumulations of nitrogen-rich ices.

Upland terrains: Fretted and eroded mantle terrain

*Fretted terrain:* Northwest of Burney crater (Fig. S1), fretted terrain consists of bright plains separated into polygons by a network of darker troughs 3 to 4 km wide (Fig. S5). Numerous craters (<25 km diameter) occur within this terrain, some with dark floors. The fretted terrain perhaps originated as tectonically disrupted blocks, whose margins were subsequently widened by glaciation and/or sublimation.

*Eroded mantle terrain:* An eroded mantle (i.e. material draping underlying topography) characterizes the high northern latitudes to the NE of SP (Fig. S3). The density of craters is low except for a few degraded examples ~50 km across. The mantle appears smooth with convex rounded edges (Fig. S6), and may be depositional. Locally the surfaces have been eroded by steep-sided pitting that reaches 3-4 km deep. The mantle has a relatively bluish tint (Fig. 1A) and is in excess of 1 km thick in places.

Western low latitude terrains

*Smooth Plains:* An extensive, low-elevation, mottled plain occurs at the western end of Innana Fossa (Fig. S1), bounded by an arcuate scarp along its southern and western borders (Fig. S7). It has a much lower crater density than surrounding regions. Despite the younger fracture system, it is significantly smoother than intercrater plains in adjacent areas. Scarp retreat appears to have been active in this region (see arrows highlighting Piri Rupes in Fig. S7).

*Bright-halo craters:* In Vega Terra, centered at 25°N, 125°E (Fig. S1), craters feature dark floors, bright rims and bright outer rim flanks, in contrast to craters elsewhere. Even within this subset of craters, there is variation in their colors and albedos (independent of sun angle), including between north- and south-facing slopes within the same crater (Fig. S8).

## Tectonics

Fig. S9 displays several examples of prominent tectonic features across the encounter hemisphere of Pluto. Faulting that appears to have experienced little degradation is concentrated to the west of Sputnik Planum. These include the subparallel normal faults of Djanggawul Fossae, which emanate from Oort crater (Fig. S9E) and extend northwards for several hundred km (Fig. S9A). The segmented grabens of Inanna and Dumuzi Fossae are ~600 km long, reach 20 km wide and 2-3 km deep, and no rift flank uplift is observed for them (Fig. S9B). In Cthulhu Regio, the normal fault scarp of Virgil Fossa is ~950 km long and reaches 3-4 km high, and cuts Elliot crater at its eastern end (Fig. S9C). Beatrice Fossae, to the south of Virgil Fossa, reaches at least 450 km long, and a poorly illuminated south-trending set of fractures is located between Beatrice Fossae and Virgil Fossa, which trend at ~45° to the larger features. A heavily degraded graben system passes across the north pole; the graben shown in Fig. S9D reaches more than 100 km long. Fewer obvious manifestations of extensional faulting are seen to the east of Sputnik

Planum, but ~550 km long Sleipnir Fossa is a prominent example, and in its southern portion is bounded on both sides by swells of Tartarus Dorsa (Fig. S9F). At the northern extent of Sleipnir Fossa, it becomes one of a set of fractures radiating from a central focus, which is similar in appearance to novae on Venus (*79*). The cause of the spatial orientation of Pluto's extensional features may be due to shell heterogeneities or other global stress-inducing mechanisms.

On Charon, Argo Chasma was viewed very obliquely along the horizon (*6*). In lower resolution images, it is seen to have an arcuate planform (Fig. S10), and so could be related to an impact basin, or alternatively, could be a continuation of the fracture/trough system that bisects Charon's encounter hemisphere.

## Impact Craters

Crater albedo patterns and morphologies vary widely across Pluto (Figs. S11 and S12). Albedo patterns include many different combinations of bright or dark floors, rim walls, floor material, and/or central peaks. Craters at northern latitudes and to the north and east of Sputnik Planum display less albedo variation and are generally bright, but may be degraded or infilled. Crater degradation is apparent in a number of forms, including tectonic disruption, infilling of various kinds, and possible sublimation and mass wasting features (Fig. S12). Some of the bright floor material in various craters appears to be a smooth infilled deposit similar in color and appearance to Sputnik Planum, implying it may be a remnant of a once more extensive deposit, or has locally accumulated in topographic lows. Much of the dark infilling material also appears smooth. Dark infilling is likely tholin deposits (*9*).

*Pluto crater statistics*

The total cumulative crater distribution on Pluto's encounter hemisphere is shown in Fig. S13A. Because of variable lighting and Pluto's active geology, this distribution does not represent a production function (one representative of the impactor size-frequency distribution). It can, however, set a lower limit on ages. As in (*6*), model ages can be assigned according to estimates of the impacting Kuiper belt object (KBO) population (12). The KBO population is estimated at large (diameter $D \gtrsim 100$ km) sizes from astronomical observations and can be extrapolated to the smaller impactor sizes that make the observable craters under a variety of plausible assumptions; numerical integrations then provide estimates of the time rate of decay of the various Kuiper belt subpopulations. In (*12*), four models were presented for the overall size-frequency of Pluto's crater-forming impactor population, which cover a broad range of possible crater densities on Pluto's surface for any specific, true terrain age. Any measured crater density on Pluto can then be compared with these values, and different "model ages" determined. Characteristics of the measured crater size-frequency distribution may make a given model age more or less plausible.

On Pluto's encounter hemisphere, the cumulative density for $D \geq 20$ km is ~$3.5\text{x}10^{-5}$ $\text{km}^{-2}$. Comparison with model predictions (Figs. 5 and 9 in (*12*)) yields ages 4 Ga or greater for all but one impact flux model. The single, younger model age (~500 million years) requires a very steep crater production function between 10- and 100-km crater diameter (i.e., many smaller impactors). Although the encounter hemisphere size-frequency distribution does steepen somewhat in this diameter range (cf. Fig. 12B), it is far from steep enough, so this model age is provisionally rejected (but see discussion of Charon crater statistics, next).

Crater densities for different Pluto terrains are shown in the relative differential, or R-plot, form (*80*), in Fig. S13B, with age predictions for the “broken power-law” KBO distribution from (*12*). The latter is probably the simplest, agreed-upon representation of the Kuiper belt population (*23*). Highly cratered regions are ancient for this flux model, whereas more lightly cratered regions indicate a wide range of crater retention ages on Pluto.

*Charon crater statistics*

Lighting precludes accurate crater counting in Charon’s higher latitudes. Nevertheless, this region does contain several craters >50-km in diameter (*D*), consistent with great age (*81, 11*). In particular, for a cumulative crater density of (1.1 ± 0.3) x $10^{-5}$ $km^{-2}$ for $D \geq 50$ km (13 such craters on ¼ of Charon’s surface, Fig. S15A), all of the model ages in (*12*, cf. *82*, *83*), which account for the range of plausible impactor size-frequency distributions in the Kuiper belt, are nominally older than 4 billion years.

The relative smoothness of the southern plains (Vulcan Planum), and the low sun angles of the images, allow accurate crater counts down to ~4-km diameter. The plains may appear more cratered than the northern terrain, but this is an artifact of lighting and viewing geometry. The cumulative crater size-frequency distribution lies below that of the north at large diameters (Fig. S15A), indicating a relatively younger age. In particular the differential slope for Vulcan Planum craters > 10-km-diameter is –2.9 ± 0.3 and the downturn to a very shallow crater size-frequency distribution at diameters $\lesssim$ 10 km appears to be real. Cumulative crater densities down to 10-km diameter agree with those reported in (*6*) from early, lossy images, considering statistical uncertainty. When comparing measured crater densities for Charon (Figs. S15A and B) with predictions as shown in the size-frequency distributions of Fig. 12 in (*12*), both as cumulative and R-plot versions (and when corrected by a factor of 4 (*82*, *83*)), all model ages for the plains but one are 4 Ga or greater. As with Pluto, we discount the young model age, but much more firmly, and conclude that Vulcan Planum as a whole likely dates back billions of years.

Of all the terrains on Pluto or Charon on which to count craters, the southern plains of Charon offer the best representation of the flux of smaller KBOs over time. The illumination and viewing geometry are ideal for discriminating crater forms, and the plains themselves appear to offer a resurfaced "clean slate" upon which to identify craters. There are no obvious gross regional (e.g., east-west) variations in the impact crater density either. When viewed as an R-plot, the overall crater size-frequency distribution for Vulcan Planum appears "flat" in the ~10-to-100–km diameter range (Fig. S15B), and is inconsistent with those proposed KBO size distributions that are "steep" (possessing a relatively great abundance of smaller impactors) in this size range. Specifically, steep implies a power law index $q \gtrsim 4$, with $q$ defined by the differential distribution $dN/dD \propto D^{-q}$. Moreover, the size-frequency distribution at smaller sizes ($\lesssim$10-km diameter) is exceptionally shallow. There is a distinct paucity of smaller craters, corresponding to a relative dearth of km-scale KBOs.

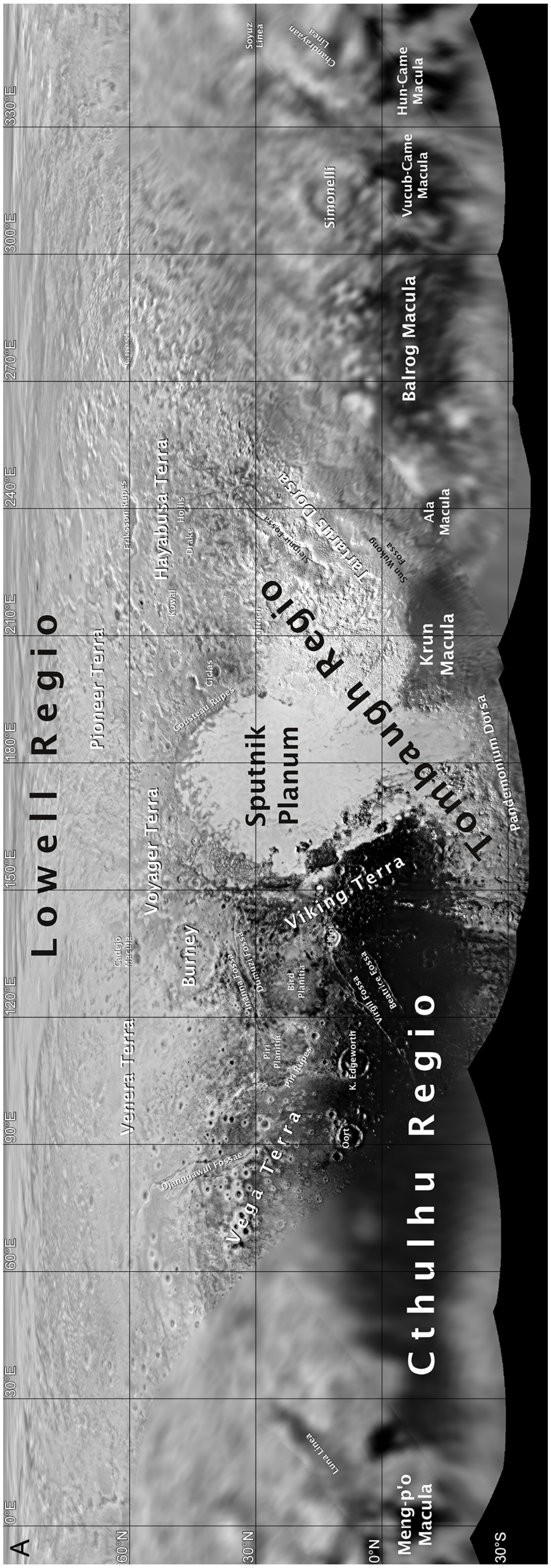

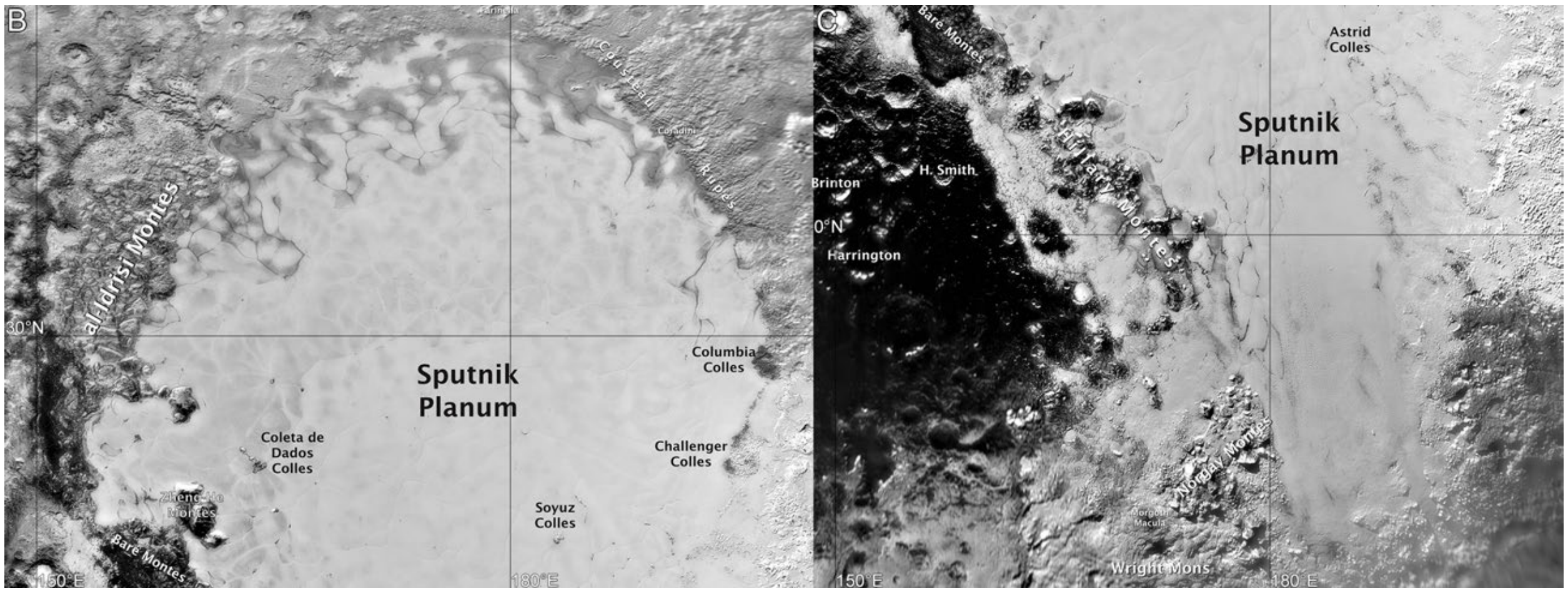

Fig. S1: Informal nomenclature maps for Pluto (A is global, B and C show details of Sputnik Planum). These maps have been updated from those in (*6*) to reflect the names in this paper.

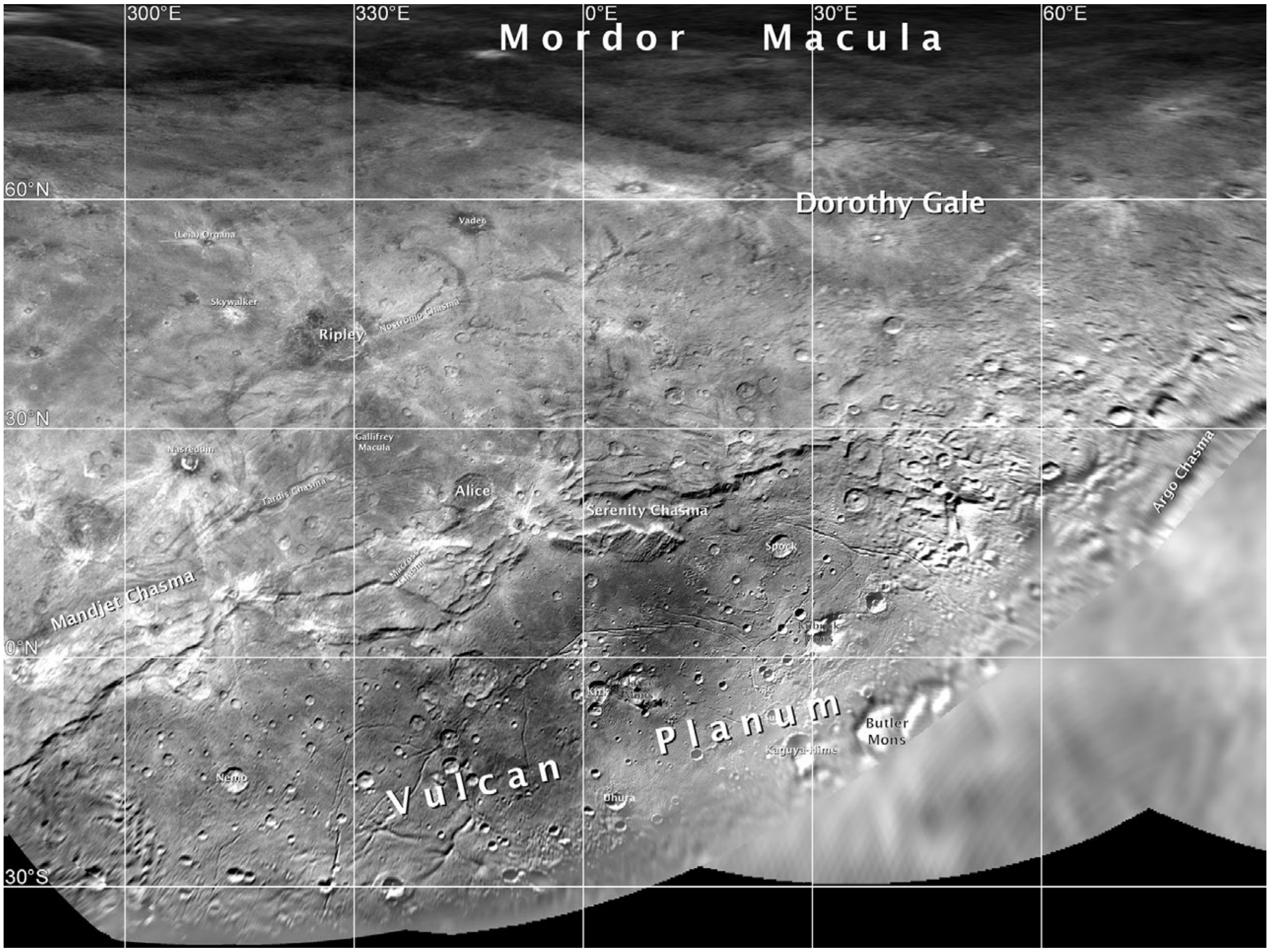

Fig. S2: Informal nomenclature map for Charon. This map has been updated from that in (*6*) to reflect the names in this paper.

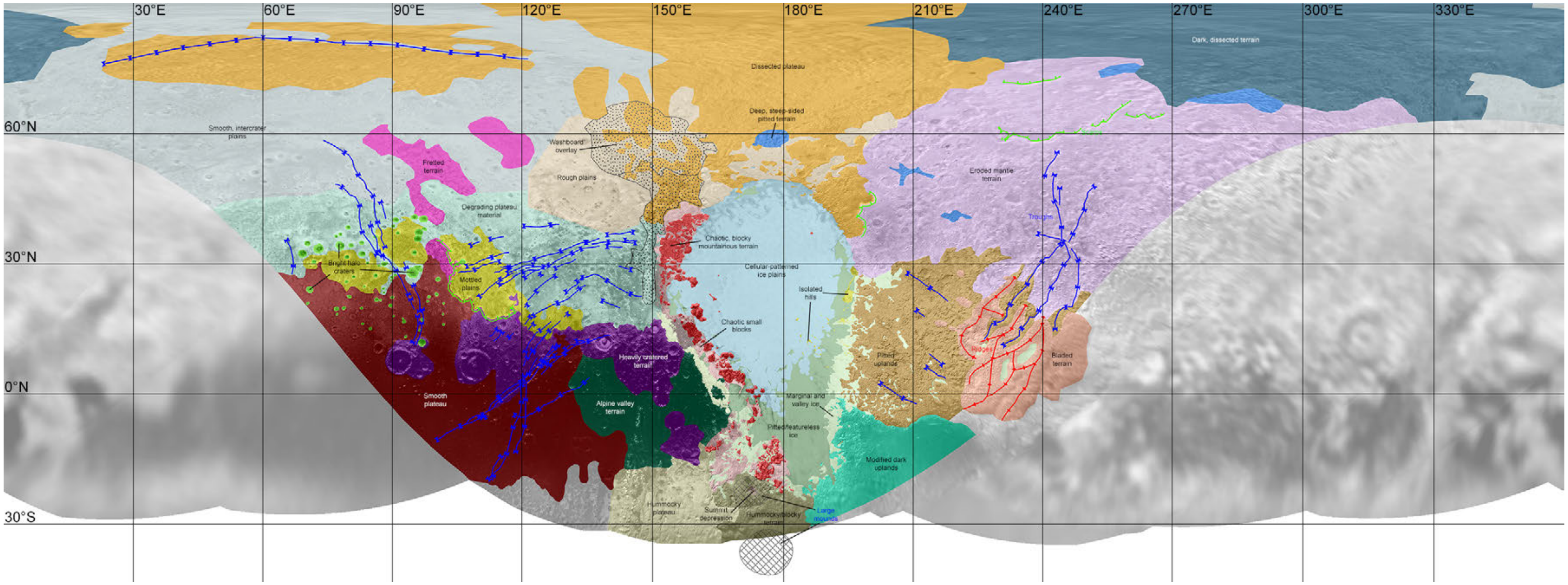

Fig. S3 (previous page): Simple cylindrical global projection of a terrain locator map for Pluto. 180° longitude is located at the center of the figure. This figure should NOT be interpreted as a geological map in any way; rather it provides a general indication of where terrains that are described in the text are located on Pluto. Terrains within low-resolution zones are not mapped.

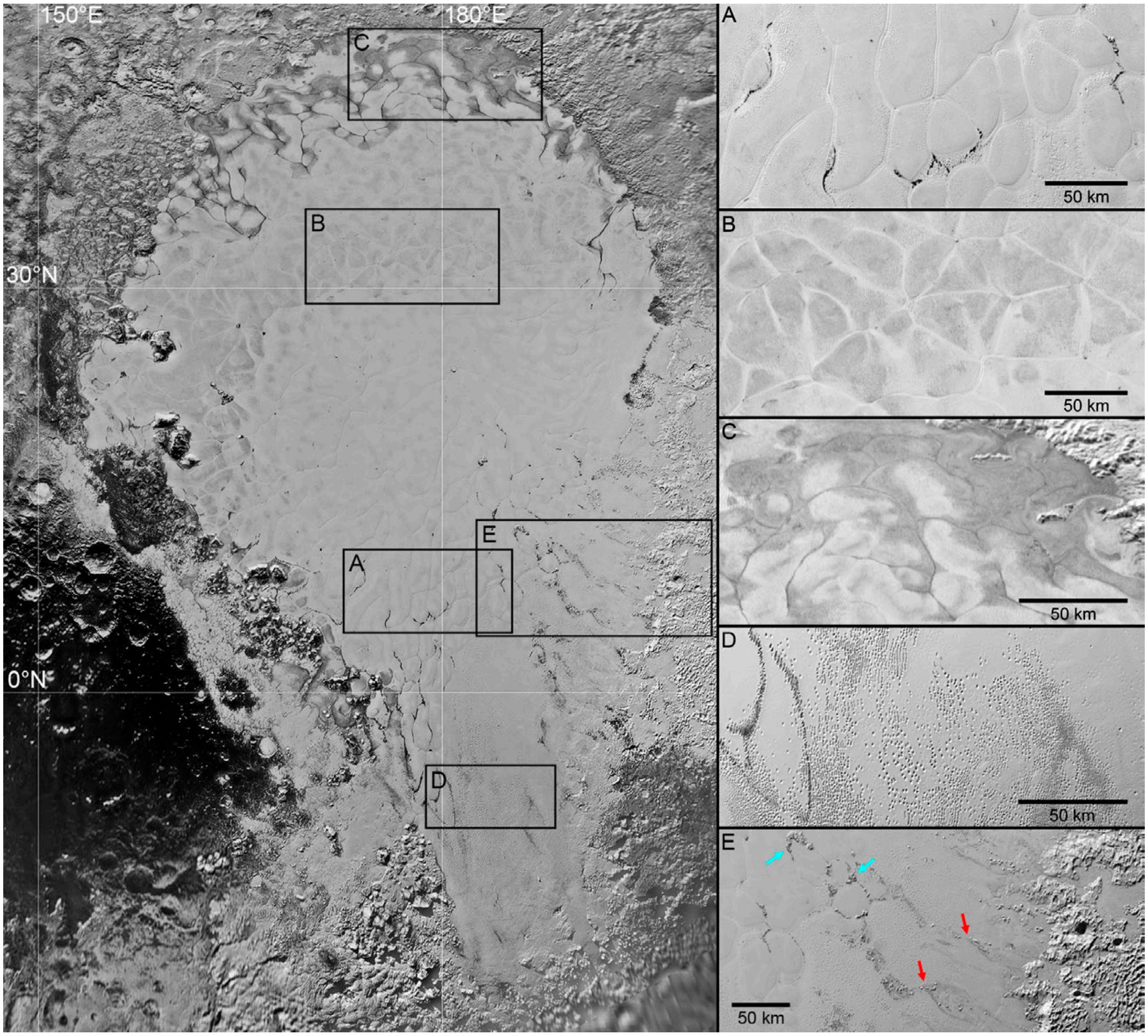

Fig. S4: **Reprojected LORRI mosaic of Sputnik Planum with details (at right) focusing on different terrain types within the plains.** Enhanced, detailed images are taken from 320 m/pixel, reprojected MVIC coverage of the P_MVIC_LORRI_CA observation. North is up in all instances. (a) Bright cellular terrain. (b) Dark cellular terrain. (c) Dark terrain displaying

lobate flow patterns at the margin with the surrounding terrain. (d) Featureless and pitted plains. (e) Isolated hills in eastern Sputnik Planum, close to the boundary with the pitted uplands (seen on the right-hand side of the figure). Chains of hills that are coincident with the paths of glaciers are highlighted by red arrows; coagulations of hills within the cellular terrain are highlighted by blue arrows.

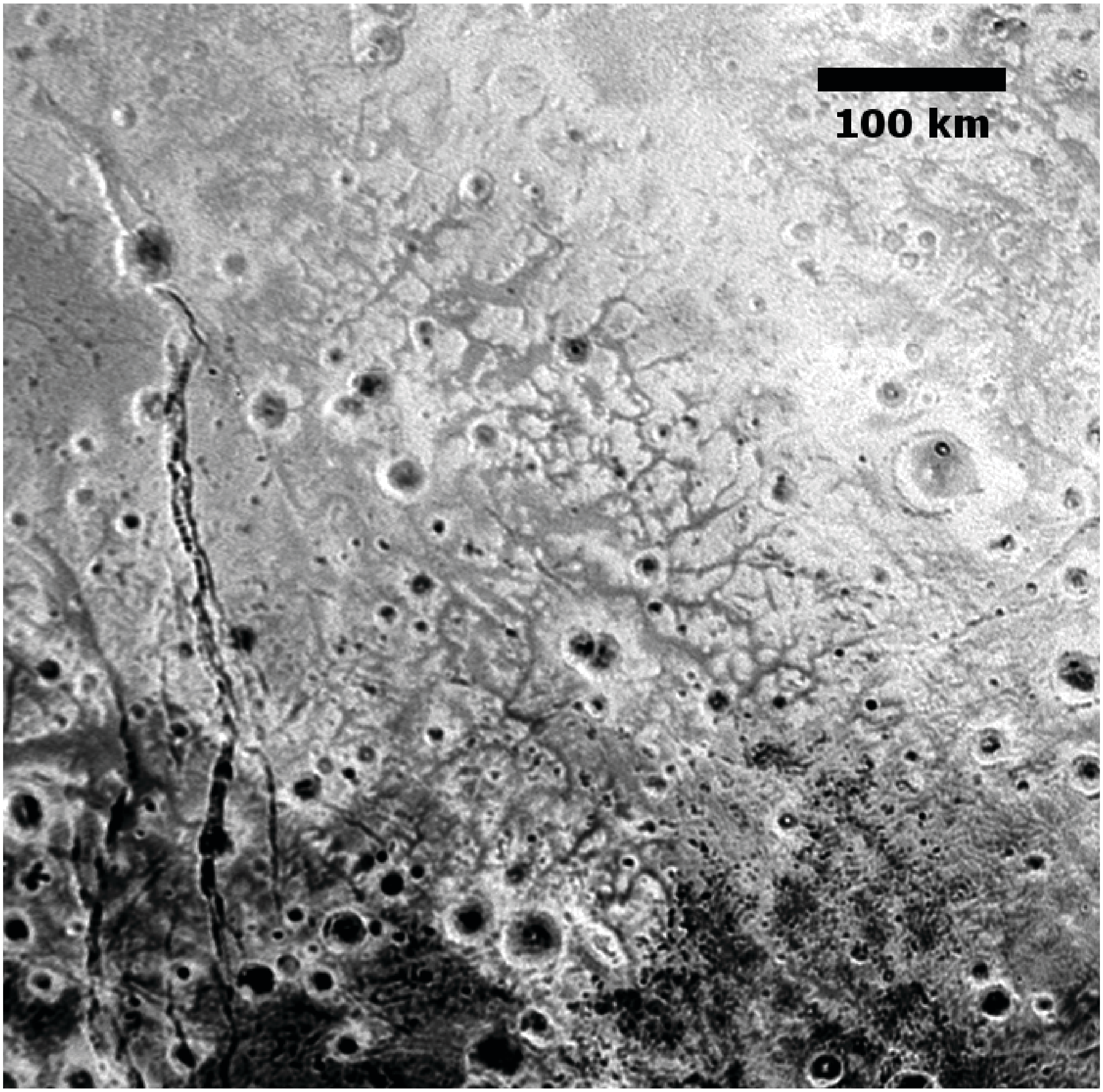

Fig. S5: **Fretted terrain located northwest of Sputnik Planum.** 890 m/pixel reprojected LORRI coverage from the P_LORRI observation, centered at 50.6°N, 97.2°E. North is up.

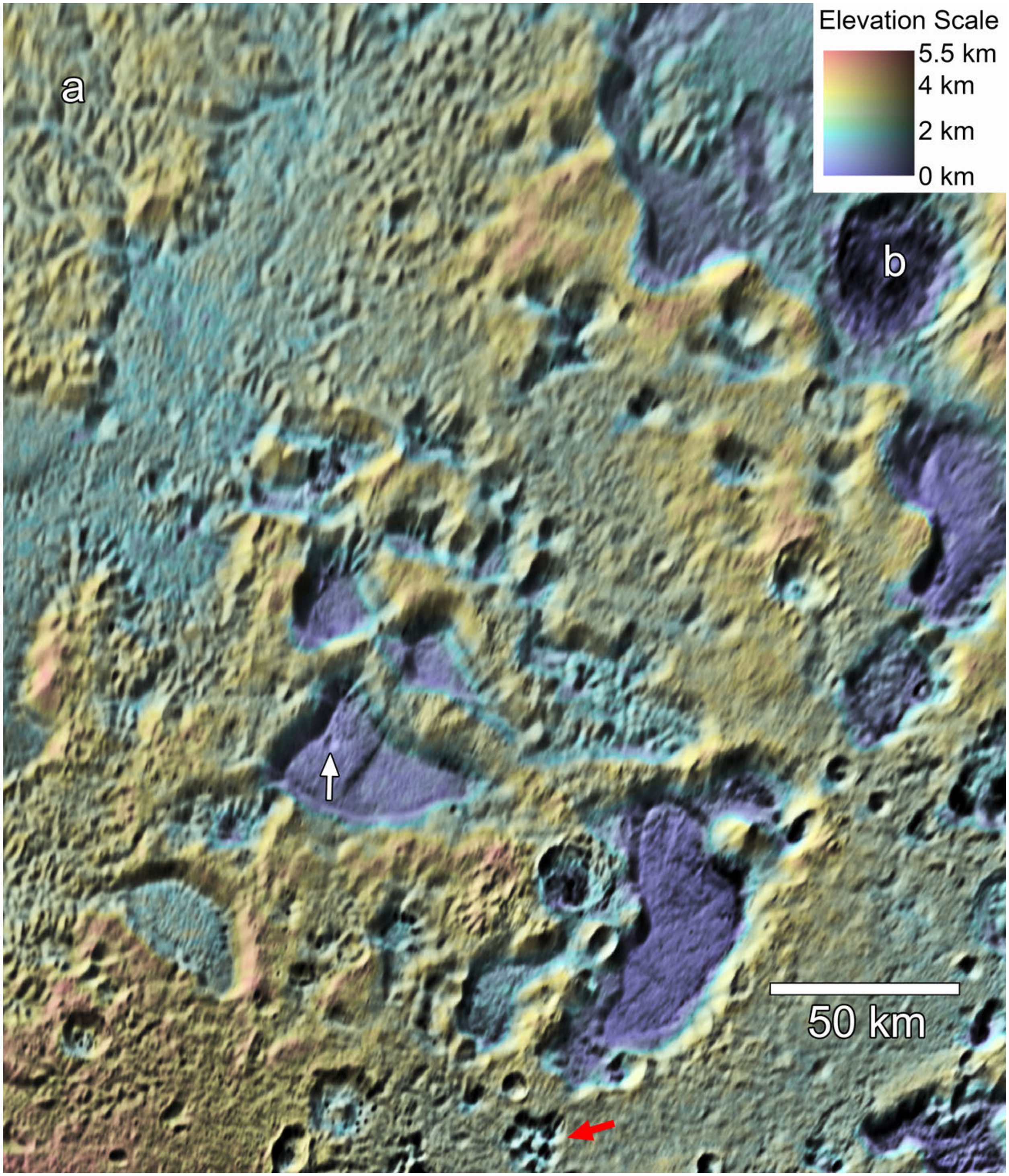

Fig. S6: **Deep, nearly flat-floored depressions in the eroded mantle terrain located northeast of Sputnik Planum.** Colorized DEM with 320 m/pixel, reprojected MVIC coverage from the P_MVIC_LORRI_CA observation overlain, centered at 59.9°N, 206.6°E. North is up. The depressions indent 1.8 to 3.5 km below the surrounding uplands. Given the numerous impact

craters on the uplands and only one small crater on the depression floors (white arrow), it is unlikely that the uplands were deposited around pre-existing depressions. Local removal of upland deposits through selective sublimation is one possibility, as are resurfacing and associated erosion by cryovolcanic flows or accumulation of ices from condensation within pre-existing depressions. The depression floors reach up to ~3 km in depth below their rims. Smaller, pit-like depressions also occur at scattered locations (red arrow). A 33 km diameter crater is located at "b" and a region of dendritic valleys occurs at "a".

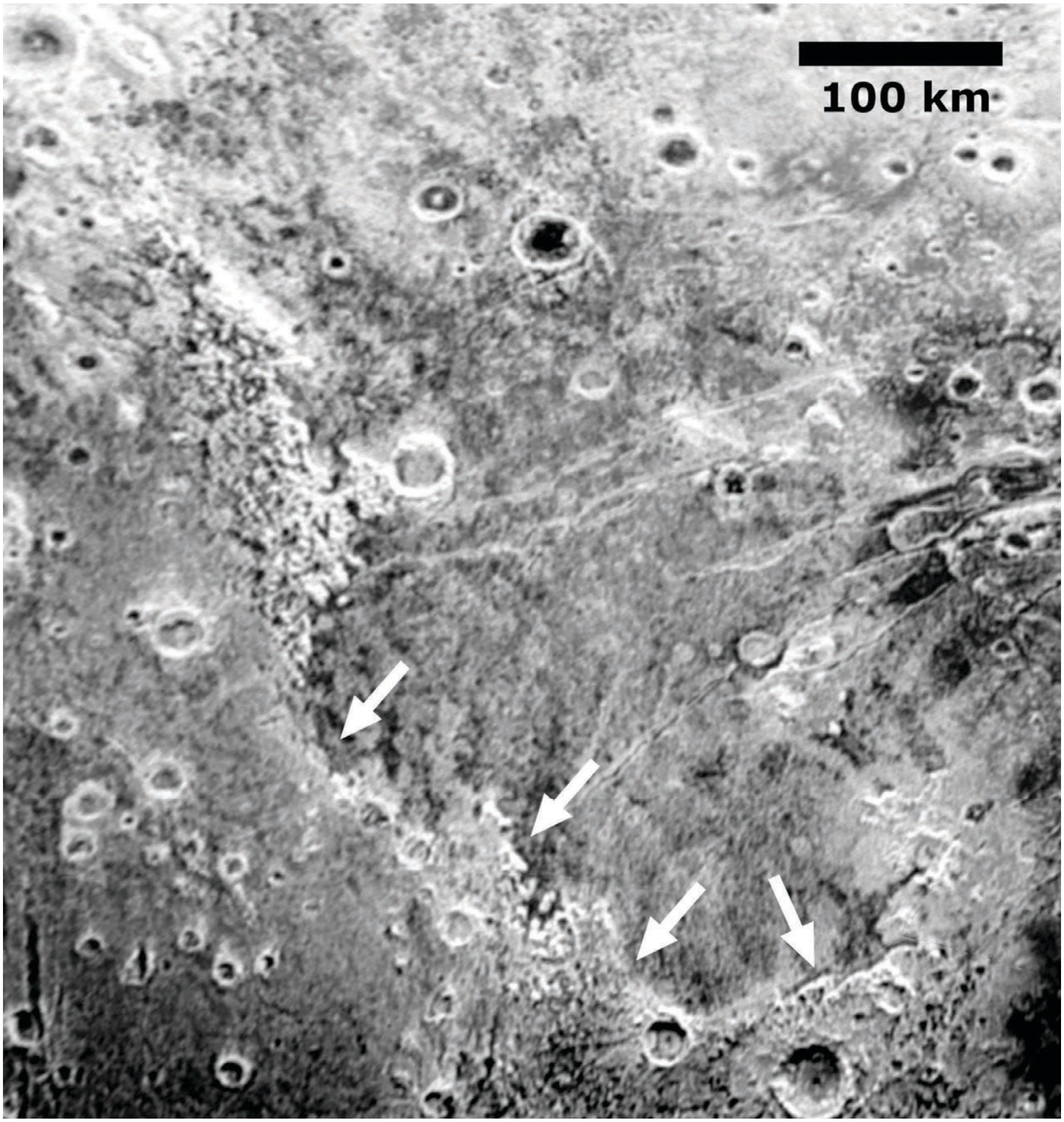

Fig. S7: **Mottled plains and a scarp located north of Cthulhu Regio.** 890 m/pixel, reprojected LORRI coverage from the P_LORRI observation, centered at 29°N, 109°E. North is up. The mottled plains in the center of the figure may have been exhumed by a scarp that surrounds them to the west and south (white arrows) that has undergone retreat. We cannot rule out the possibility that these plains may have also experienced subsequent resurfacing. Fractures from Innana Fossae cross these plains from the east.

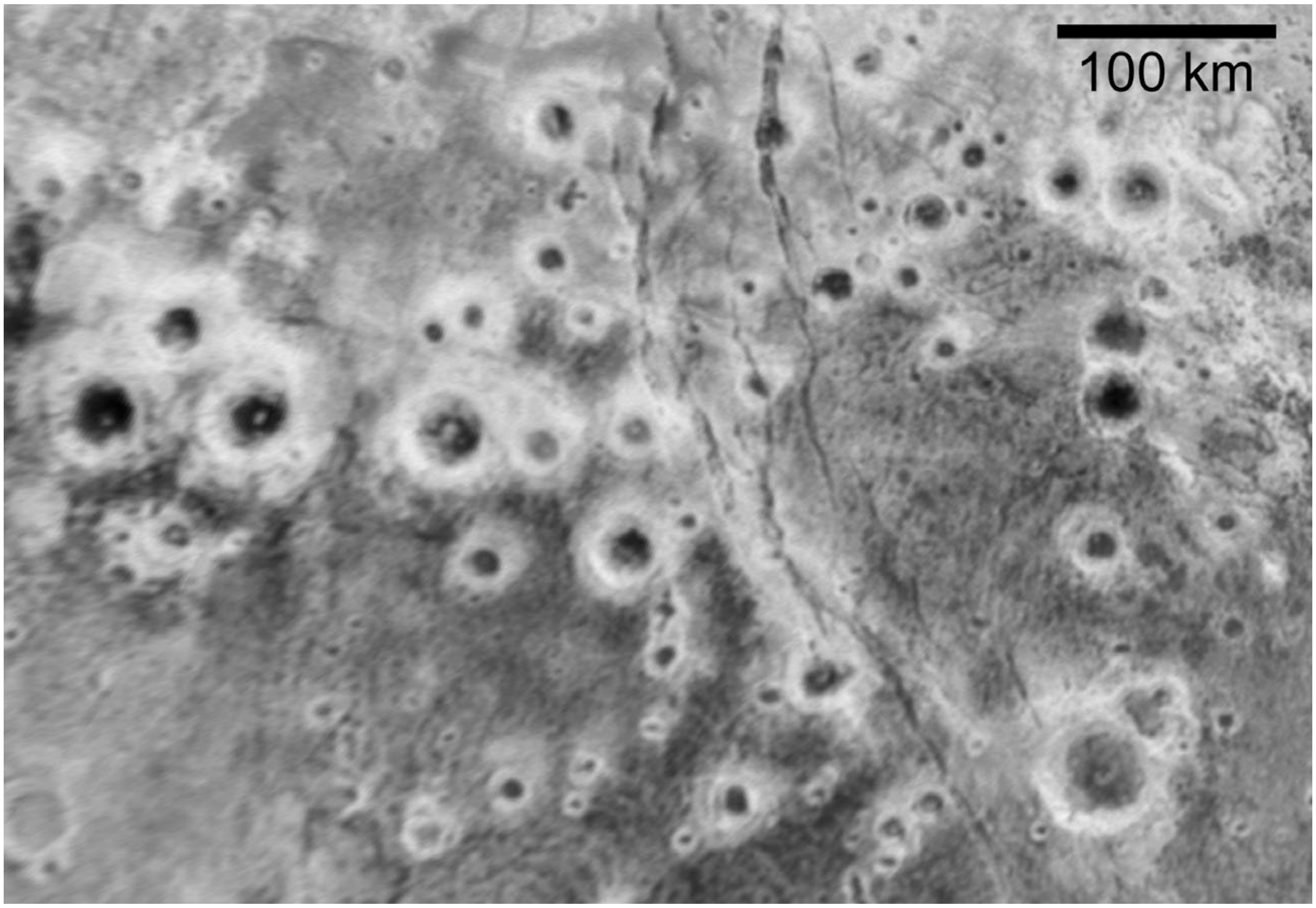

Fig. S8: **Bright-halo craters located in Vega Terra.** Djanggawul Fossae runs N-S through the center of the image. 890 m/pixel, reprojected LORRI coverage from the P_LORRI observation, centered at 34.3°N, 83.2°E. North is up.

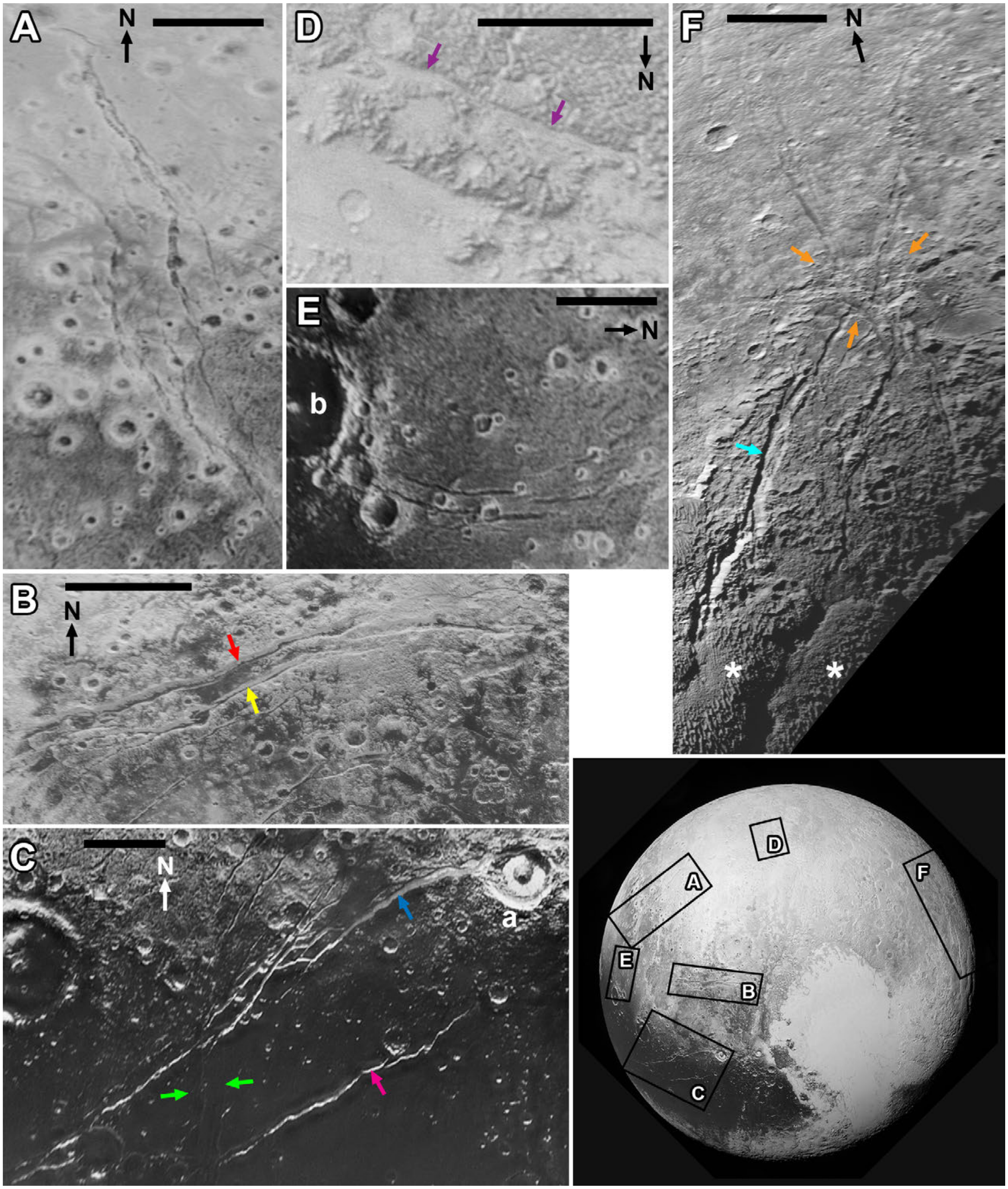

Fig. S9: **Tectonic features on Pluto.** Locations of the features on are highlighted in the figure at bottom-right. A and C-E are taken from 890 m/pixel, reprojected LORRI coverage from the P_LORRI observation, B is taken from 320 m/pixel, reprojected MVIC coverage of the P_MVIC_LORRI_CA observation, and F is taken from 680 m/pixel MVIC coverage of the P_COLOR2 observation. Scale bars all measure 100 km. (A) Djanggawul Fossae; image centered at 41.1°N, 83.8°E. (B) Inanna (red arrow) and Dumuzi (yellow arrow) Fossae; image centered at 30.9°N, 131.5°E. (C) Virgil (blue arrow) and Beatrice (pink arrow) Fossae and Elliot crater (a), with the green arrows indicating a poorly illuminated south-trending set of fractures; image centered at 6.8°N, 126.1°E. (D) A heavily degraded graben in the northern uplands (purple arrows); image centered at 81.2°N, 99.1°E. (E) Djanggawul Fossae, emanating from Oort crater (b); image centered at 16.8°N, 92.3°E. (F) A set of fractures radiating from a central focus (orange arrows) to the north of Tartarus Dorsa (asterisks) and Sleipnir Fossa (cyan arrow); image centered at 31.8°N, 243.5°E.

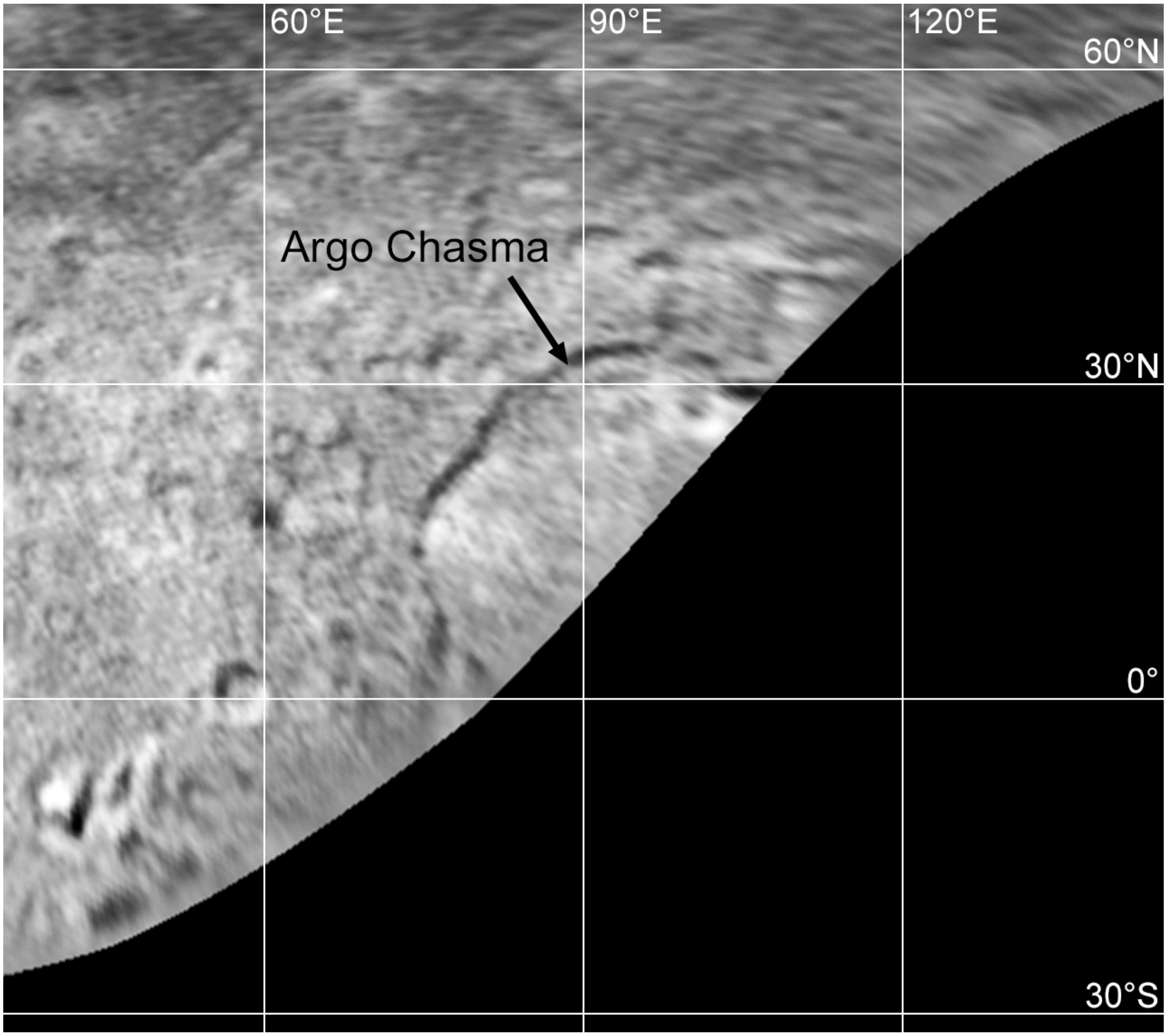

Fig. S10: **Argo Chasma in plan view.** Simple cylindrical projection of a portion of 5.32 km/pixel LORRI coverage from the PCNH_MULTI_LONG_1d1_01 observation. Projection covers Charon longitudes from 0° to 180°. Argo Chasma, seen obliquely during the encounter (Fig. S3 in (*6*)), is resolved in plan to be arcuate over at least 180°, and with a radius of ~175 km. Argo could thus be the rim of a (relaxed) impact basin, or alternatively, be tectonic in origin, and possibly related to the encounter hemisphere chasmata.

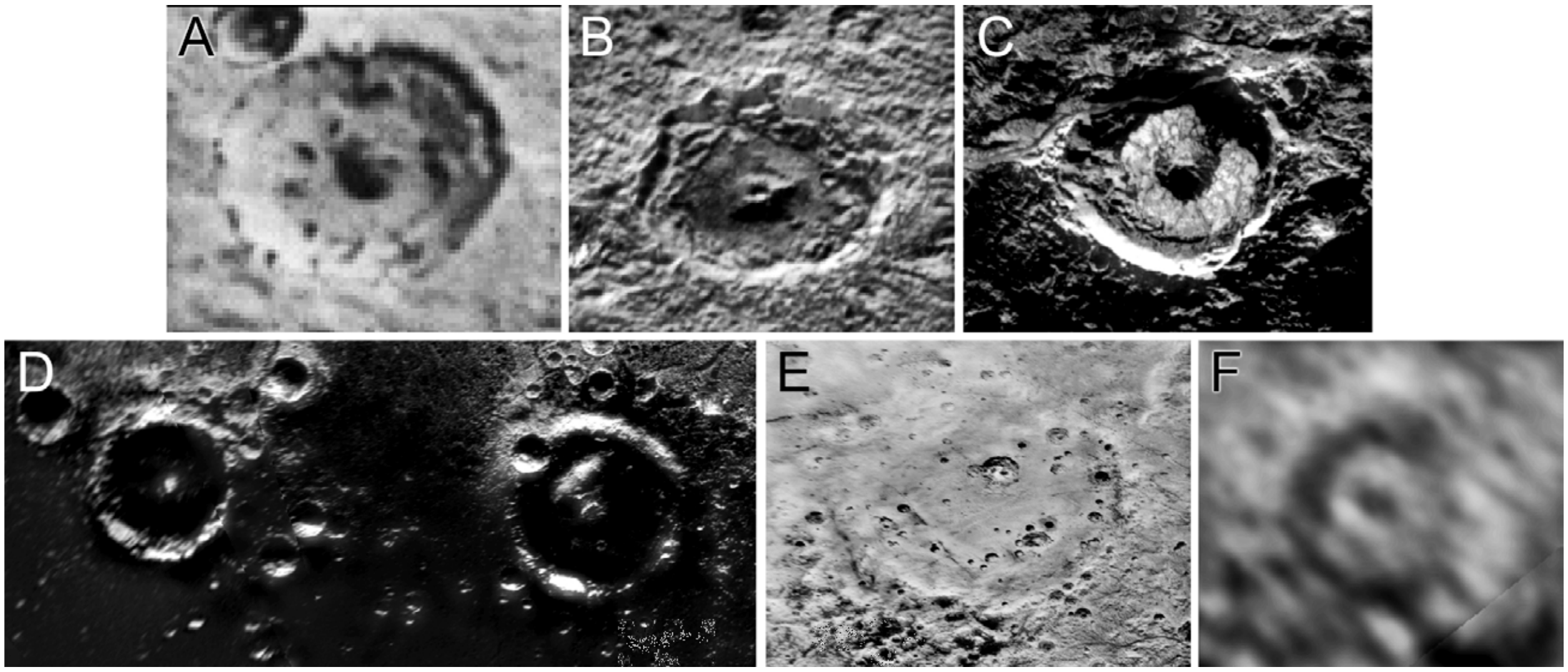

Fig. S11: **A variety of crater morphologies and albedos seen across Pluto.** (a) is 125 m/pixel LORRI coverage from the P_MPAN_1 observation; (b) to (e) is 320 m/pixel, reprojected MVIC coverage from the P_MVIC_LORRI_CA observation; (f) is 12.5 km/pixel LORRI coverage from the NAV_C4_L1_CRIT_37_01 observation. North is up in all cases. (a) 17 km diameter crater at 39°N, 139.5°E; (b) 45 km diameter Giclas crater at 39.5°N, 201.5°E; (c) 88 km diameter Elliot crater at 12°N, 138.5°E; (d) 120 km diameter Oort crater at 7.5°N, 91.5°E (left), and 140 km diameter Edgeworth crater at 6.5°N, 108.5°E; (e) 250 km diameter Burney crater at 45°N, 134.5°E; (f) 250 km diameter Simonelli crater at 13°N, 313.5°E.

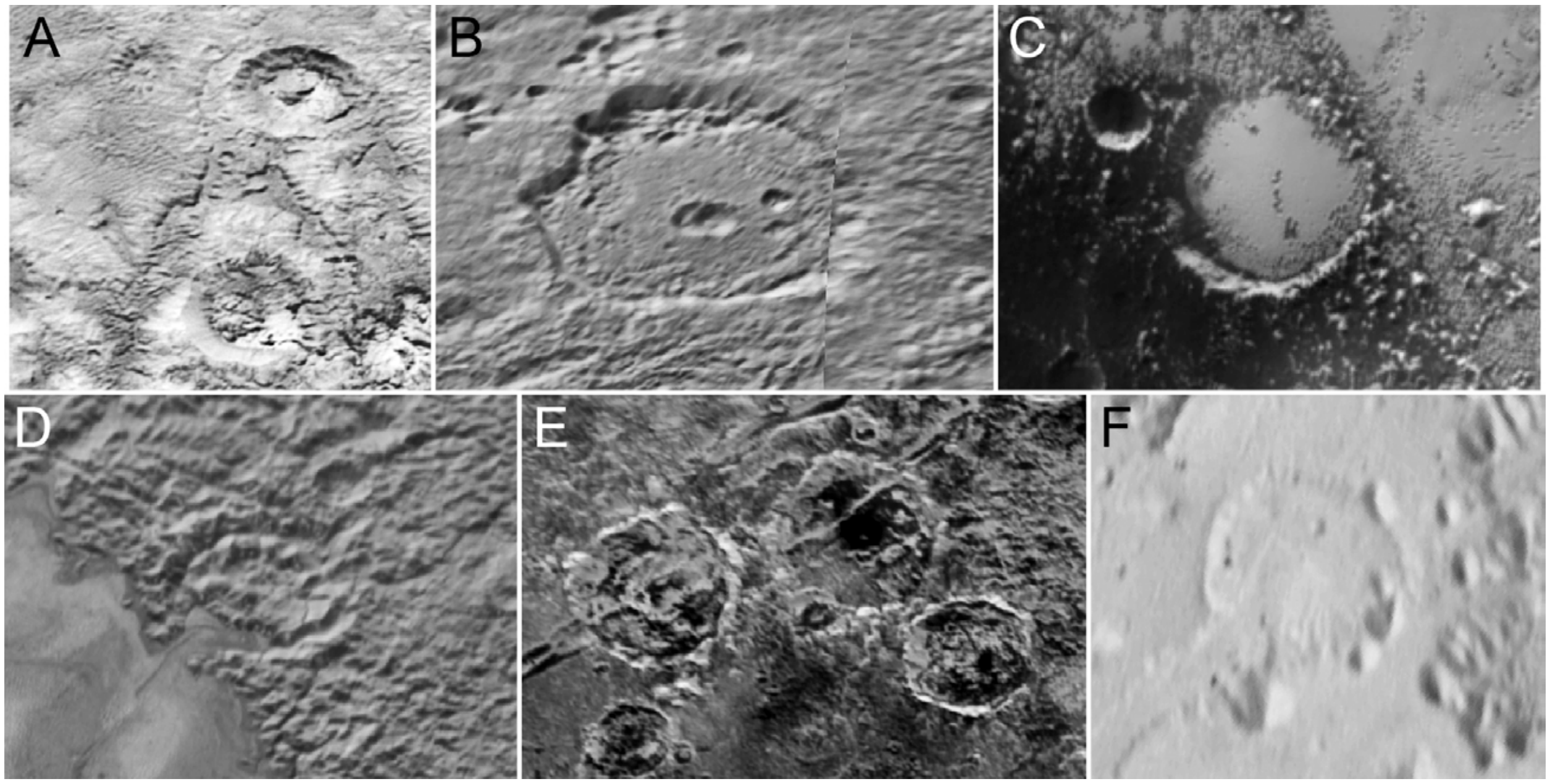

Fig. S12: **A variety of forms of crater degradation seen across Pluto.** (a) is from 390 m/pixel, reprojected LORRI coverage from the P_LORRI_STEREO_MOSAIC observation; (b) through (f) are from 320 m/pixel, reprojected MVIC coverage from the P_MVIC_LORRI_CA observation. (a) Craters displaying fluted walls. The two larger craters are each ~35 km across. Image centered at 46.5°N, 151°E. (b) 60 km diameter Kowal crater at 49°N, 217.5°E in the eroded mantle terrain. (c) 32.5 km diameter crater on the margin of Sputnik Planum at 4°S, 167°E, with ice covering its floor. (d) 36 km diameter Coradini crater on the margin of Sputnik Planum at 42.5°N, 191.5°E, which appears to have been dissected by steep, branched valley networks. (e) Craters cut by extensional fracturing in Viking Terra west of Sputnik Planum. The two larger craters are each ~45 km across. Image centered at 21.5°N, 130°E. (f) 37 km diameter crater at 78°N, 136°E, with a degraded rim and which appears to have been mantled by a bright deposit.

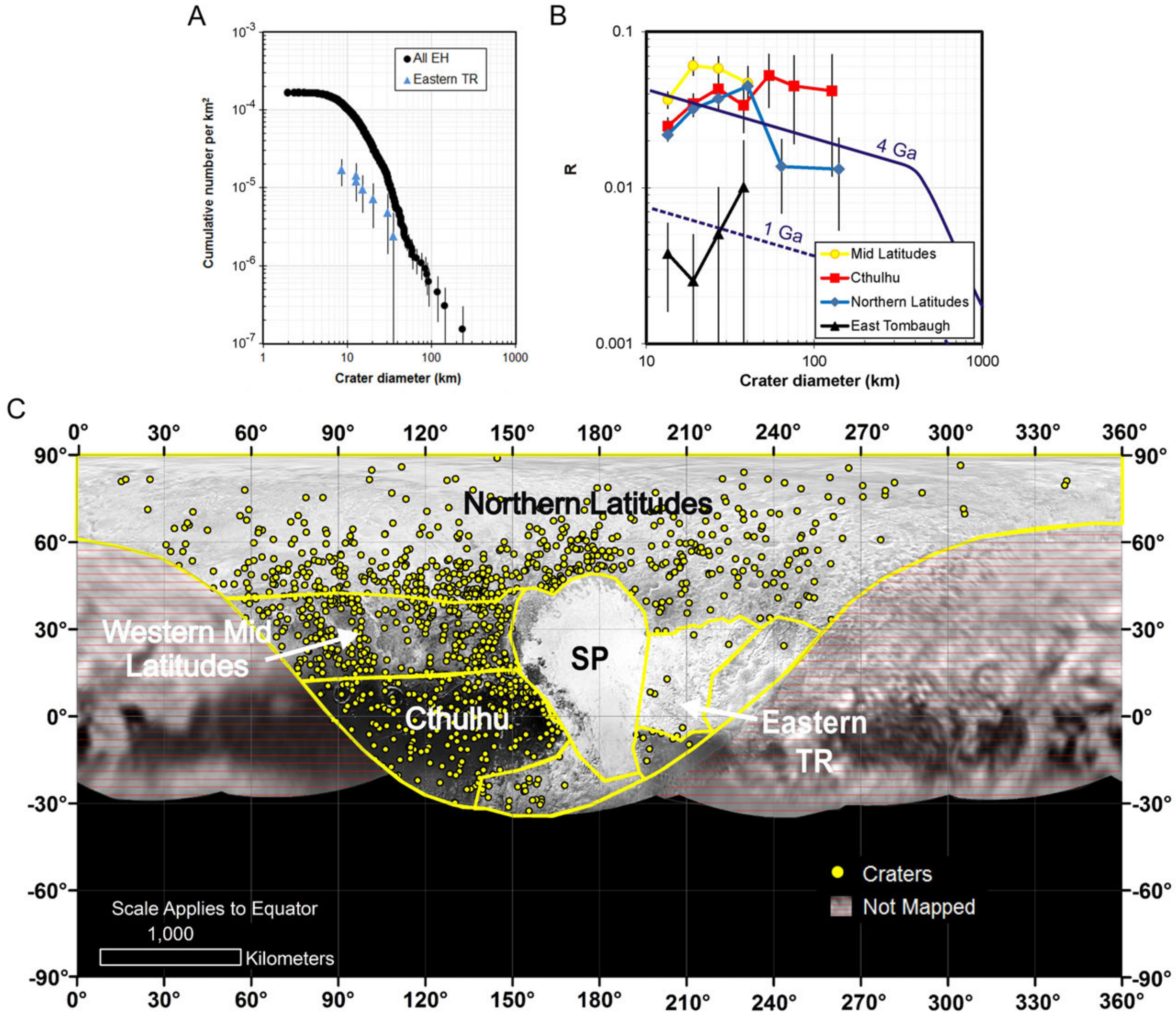

Fig. S13: **Crater statistics for Pluto.** (A) Cumulative size-frequency distribution for all craters in Pluto's encounter hemisphere, and those just within eastern Tombaugh Regio (TR), showing higher crater densities and an older age for Pluto as a whole compared with eastern Tombaugh Regio. (B) R-plot for craters within specific terrains in Pluto's encounter hemisphere. R-plots (or 'Relative' plots) plot differential crater counts relative to (divided by) a -3 power law distribution ($dN = kD^{-3}$), and highlight differences between different diameter ranges and surface units (*80*). The 4 Ga and 1 Ga age lines are predictions for expected crater densities for a given surface age for the broken power-law, or "knee", impactor distribution in (*12*), for reference; other impactor distributions in (*12*) give higher or lower R values for the same model age. (C) Map of the locations of all craters counted in Pluto's encounter hemisphere. Craters were mapped using 890 m/pixel LORRI coverage from the P_LORRI observation. Poisson ($\sqrt{N}$) errors are assumed in A and B.

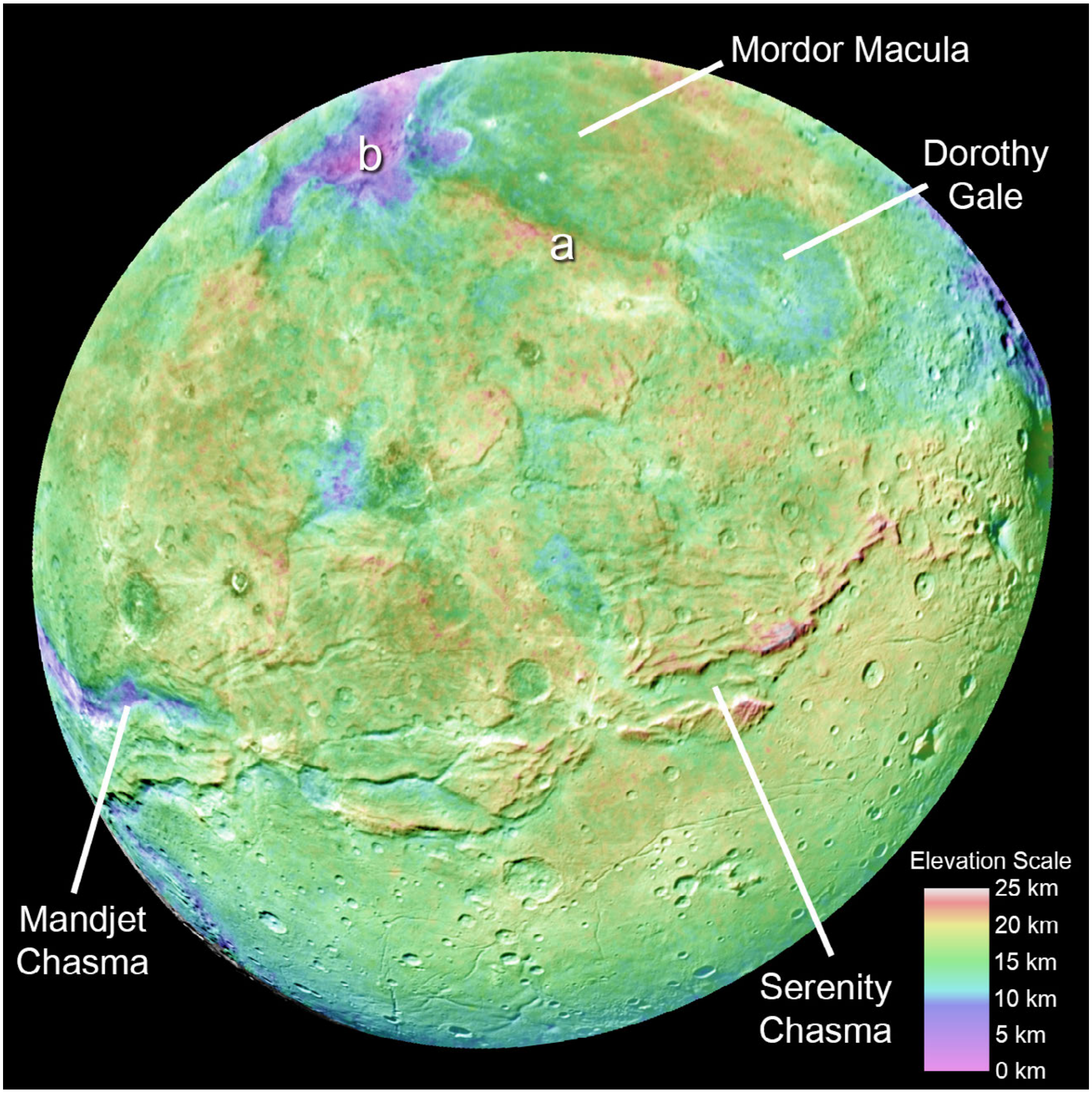

Fig. S14: **Colorized DEM of Charon's encounter hemisphere**. DEM is overlain on 1460 m/pixel MVIC coverage of the C_COLOR_2 observation. North is up. A prominent ridge (a) and an irregular depression (b) in the north polar region are highlighted.

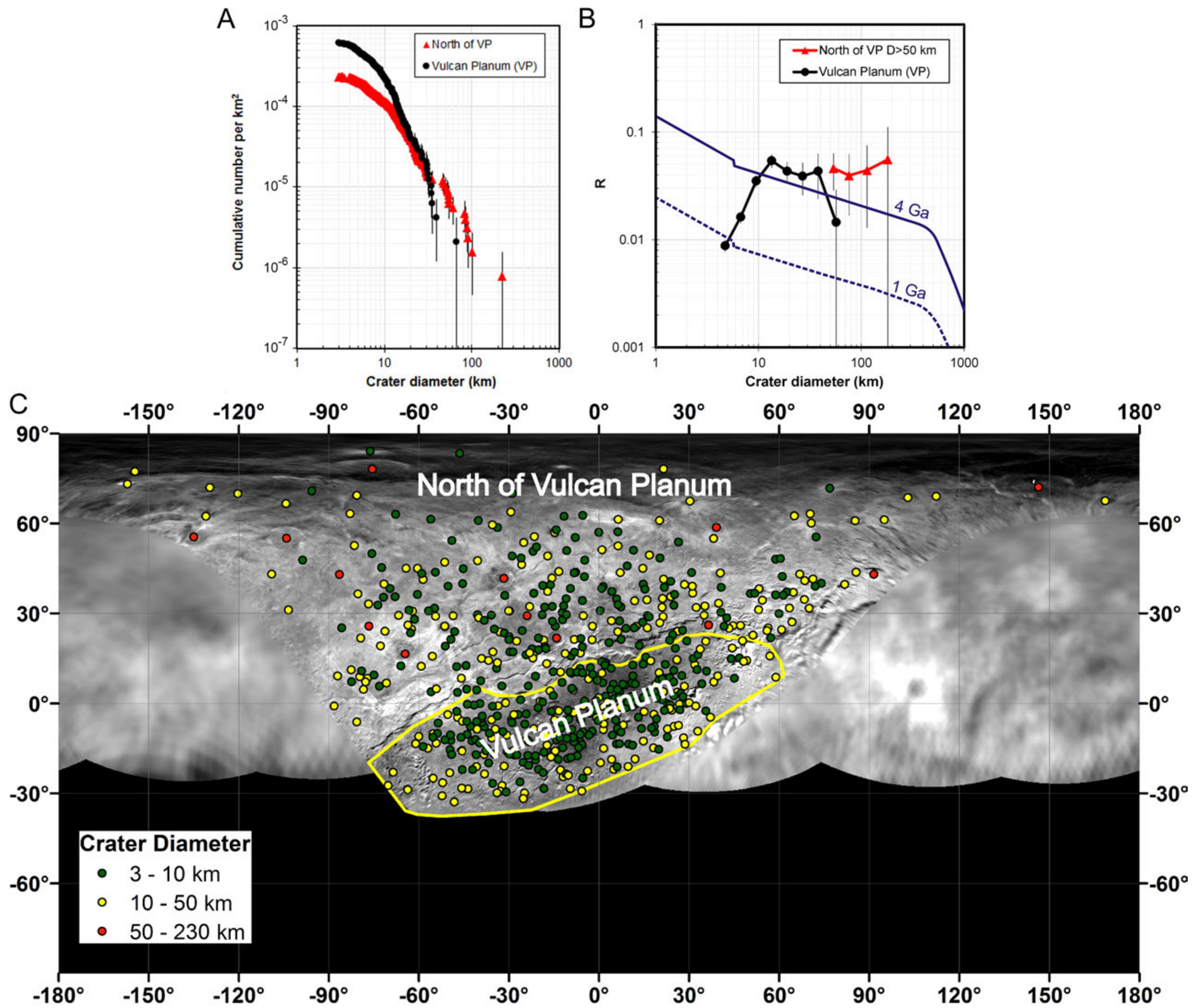

Fig. S15: **Crater statistics for Charon.** (A) Cumulative size-frequency distributions for craters within Vulcan Planum (VP) and craters north of Vulcan Planum on Charon. (B) R-plot for craters within Vulcan Planum and craters north of Vulcan Planum. (C) Map of the locations of all craters counted in Charon's encounter hemisphere. Craters were mapped using 890 m/pixel LORRI coverage from the C_LORRI observation. Poisson ($\sqrt{N}$) errors are assumed in A and B.

**Table S1. Possible convective properties of a 4-km deep $N_2$ ice layer on Pluto.**

| $\delta T/h/\Delta T$ | 3 K/100 m/20 K | 3 K/1 km/20 K | 3 K/100 m/9 K | 6 K/100 m/9 K |
|---|---|---|---|---|
| $\Delta\eta$ | 800 | 800 | 20 | 4.5 |
| $\sigma$ (Pa) | 40 | 400 | 40 | 75 |
| $T_b$ (K) | 56 | 56 | 45 | 45 |
| $\eta_b$ (Pa-s) | $4.5\times10^{12}$ | $2.5\times10^{11}$ | $2.5\times10^{13}$ | $1\times10^{13}$ |
| $Ra_b$ | $2.5\times10^{6}$ | $5\times10^{7}$ | $2\times10^{5}$ | $5\times10^{5}$ |
| $Nu$ | 7 | 18 | 8 | 16 |
| $q$ (mW/m$^2$) | 7 | 18 | 3.5 | 7.5 |

Temperature contrast ($\delta T$) and corresponding length ($h$) scale within the convective flow determines differential stress level ($\sigma$), which in turn determines basal viscosity ($\eta_b$) and Rayleigh number ($Ra_b$), and viscosity ratio $\Delta\eta = exp(\Delta T/\delta T)$ across the layer for a chosen basal temperature ($T_b$) and total temperature difference $\Delta T$ (from top to bottom). Different example combinations of $\delta T$, $h$, and $\Delta T$ are shown. Dimensionless and dimensional heat flows ($Nu$ and $q$) are then estimated from scaling (*48*).